\documentclass[prx,aps,twocolumn,reprint,floatfix,showpacs,superscriptaddress]{revtex4-1}

\usepackage{xfrac}
\usepackage{booktabs}
\usepackage{array}

\usepackage{mathrsfs}
\usepackage{latexsym}
\usepackage{amssymb}
\usepackage{amsmath}
\usepackage{amsfonts}
\usepackage{wasysym}

\usepackage{microtype}
\usepackage[varg]{txfonts}
\usepackage[percent]{overpic}

\usepackage{braket}
\usepackage{titlesec}
\usepackage{comment}
\usepackage{verbatim}

\usepackage{array}
\newcommand{\PreserveBackslash}[1]{\let\temp=\\#1\let\\=\temp}
\newcolumntype{C}[1]{>{\PreserveBackslash\centering}p{#1}}
\newcolumntype{R}[1]{>{\PreserveBackslash\raggedleft}p{#1}}
\newcolumntype{L}[1]{>{\PreserveBackslash\raggedright}p{#1}}

\usepackage{color}
\usepackage[colorlinks,bookmarks=false,citecolor=blue,linkcolor=blue,urlcolor=blue,pdfstartview=FitH]{hyperref}

\definecolor{darkred}{rgb}{0.7,0.0,0.0}

\definecolor{darkblue}{rgb}{0,0.02,0.45}

\definecolor{darkgreen}{rgb}{0.02,0.45,0.0}

\definecolor{violet}{rgb}{0.8,0.2,0.6}

\usepackage{graphicx}
\usepackage{subfigure}
\usepackage{pict2e}
\usepackage{multirow}

\newcommand{\be}{\begin{equation}}
\newcommand{\ee}{\end{equation}}
\newcommand{\bea}{\begin{eqnarray}}
\newcommand{\eea}{\end{eqnarray}}
\newcommand{\sbe}{\small\begin{equation}}
\newcommand{\see}{\end{equation}\normalsize}
\newcommand{\sbea}{\small\begin{eqnarray}}
\newcommand{\seea}{\end{eqnarray}\normalsize}

\graphicspath{{Figures/}}

\def\bs{\boldsymbol}
\def\vec{\mathbf}
\def\mc{\mathcal}

\usepackage{times}

\begin{document}

\title{Quantum-classical crossover in the spin-1/2 Heisenberg-Kitaev kagome magnet}

\author{Yang Yang}
\affiliation{School of Physics and Astronomy, University of Minnesota, Minneapolis, MN 55455, USA}

\author{Natalia B. Perkins}
\affiliation{School of Physics and Astronomy, University of Minnesota, Minneapolis, MN 55455, USA}

\author{Fulya Ko\c{c}}
\affiliation{School of Physics and Astronomy, University of Minnesota, Minneapolis, MN 55455, USA}

\author{Chi-Huei Lin}
\affiliation{School of Physics and Astronomy, University of Minnesota, Minneapolis, MN 55455, USA}

\author{Ioannis Rousochatzakis}
\affiliation{Department of Physics, Loughborough University, Loughborough LE11 3TU, United Kingdom}

\date{\today}

\begin{abstract}
The spin-1/2 Heisenberg kagome antiferromagnet is one of the paradigmatic playgrounds for frustrated quantum magnetism, with an extensive number of competing resonating valence bond (RVB) states emerging at low energies,  including gapped and gapless spin liquids and valence bond crystals.  
Here we revisit the crossover from this quantum  RVB phase to a semiclassical regime brought about by anisotropic Kitaev interactions, and focus on the precise mechanisms underpinning this crossover. 
To this end, we introduce a simple parametrization of the classical ground states (GSs) in terms of emergent Ising-like variables, and use this parametrizaton: i) to construct an effective low-energy description of the order-by-disorder mechanism operating in a large part of the phase diagram, and ii) to contrast, side by side, exact diagonalization data obtained from the full basis with that obtained from the restricted (orthonormalized) basis of classical GSs.
The results reveal that fluctuation corrections from states outside the restricted basis are strongly quenched inside the semiclassical regime (due to the large anisotropy spin gaps), and that the RVB phase survives up to a relatively large value of Kitaev anisotropy $K$. 
We further find that the pure Kitaev model admits a subextensive number of one-dimensional symmetries, which explains naturally the absence of classical and quantum order by disorder reported previously.
\end{abstract}

\maketitle

\vspace*{-1cm}
\section{Introduction}
\vspace*{-0.3cm}
In recent years, quantum materials with strong spin-orbit coupled 4d and 5d ions and bond-dependent anisotropic interactions have been the subject of much experimental and theoretical work~\cite{KimKim2008,Jackeli2009,BookCao,Rau2016,Winter2017,Trebst2017,Knolle2017,Takagi2019,Motome2019}.
Unlike isotropic Heisenberg magnets, these materials break explicitly the SU(2) spin rotational invariance down to a discrete subgroup which is set by the interplay of spin-orbit coupling, crystal field effects and electronic correlations. The resulting anisotropic exchange gives rise to a new type of magnetic frustration, different from geometrical frustration~\cite{Diep2005,HFMBook}, a wealth of unusual magnetic orders with strong sensitivity to external perturbations~\cite{Trebst2017,Chern2017,Kasahara2018,Janssen2019,Takagi2019,Gordon2019,Rousochatzakis2018,Li2019,Li2020,KimBaek2020,LeeKimBaek2020}, as well as gapped and gapless spin liquids with fractionalized excitations~\cite{Kitaev2006}. 
In addition to the extensively studied layered honeycomb materials $\alpha$-RuCl$_3$, Na$_2$IrO$_3$ and $\alpha$-Ir$_2$IrO$_3$, and their 3D analogues ($\beta$-$\gamma$)-Li$_2$IrO$_3$, other geometries -- including triangular, kagome, pyrochlore, hyperkagome and fcc lattices -- have attracted a lot of attention because they combine the frustration from the competing exchange couplings with the geometric frustration of the underlying lattices~\cite{Kimchi2014,Jackeli2015,Becker2015,Catuneanu2015,Rousochatzakis2016,Mengqun2019,Li2017a,Ducatman2018,Kishimoto2018,Morita2018,Morita2019b,Morita2019a}.

Here we revisit the spin-1/2 Heisenberg-Kitaev (or $JK$-) model on the kagome lattice, which may be relevant for some rare-earth based compounds of the type A$_2$RE$_3$Sb$_3$O$_{14}$, where A=Mg or Zn and RE is a rare-earth ion~\cite{Scheie2016,Dun2016,Dun2017,Paddison2016,Sanders2016}.
This model interpolates between the spin-1/2 kagome Heisenberg antiferromagnet (KHAF) -- 
one of the paradigmatic playgrounds of competing resonating valence bond (RVB) states, including valence bond crystals as well as gapped and gapless spin liquids~\cite{HFMBook,Balents2010,Sachdev1992,
ZengElser95,Lecheminant1997,Mila1998,SinghHuse07,Sindzingre2009,Poilblanc2010,Evenbly2010,YanHuseWhite2011,Gotze2011,Shollwock2012,Iqbal2013,Capponi2013,Xie2014,Rousochatzakis2014,He2017,Liao2017,Mei2017,Ralko2018,Laeuchli2019} 
-- 
and the compass-like, Kitaev model in which the coupling in spin space is tied to the orientation of the bonds~\cite{Nussinov2015}.

The $JK$ model on the kagome lattice has been studied previously by Kimchi and Vishwanath~\cite{Kimchi2014} and by Morita {\it et al}~\cite{Morita2018,Morita2019b}, and a lot of results are already known, including most of the aspects of the classical ground state (GS) phase diagram~\cite{Kimchi2014,Morita2018}, a numerical demonstration of an order by disorder mechanism operating in a large part of the parameter space, and the absence of this mechanism in the pure Kitaev model~\cite{Morita2018}. However, the microscopic origin of the order by disorder mechanism and its absence in the pure Kitaev model has not been understood.  More importantly, the question of whether the RVB phase of the KHAF remains robust in an extended parameter space (and, in particular, whether the answer depends on the nature of the ground state of the KHAF or the presence of a gap) has not been settled.

The main results from our study can be summarized as follows. First, we provide a simple parametrization of the classical GS manifold in terms of emergent Ising-like variables. This parametrization offers a convenient platform for the analysis of the quantum model. Second, we show that the pure Kitaev model admits a subextensive number of one-dimensional symmetries, which explain naturally the reported absence of classical and quantum order by disorder~\cite{Morita2018,Morita2019b}. 
Third, we perform a semiclassical perturbative expansion and derive an effective Hamiltonian in terms of the emergent Ising variables. This Hamiltonian provides a simple picture for the order by disorder effect observed  numerically by Morita {\it et al}~\cite{Morita2018}.
Fourth, we demonstrate explicitly the quantum-classical crossover between the RVB physics of the KHAF and the regime stabilized by Kitaev anisotropy. This is achieved by contrasting, side by side, ED results in the full basis and in the restricted orthonormalized basis of classical GSs. The comparison shows that the regime stabilized by the Kitaev coupling has a robust semiclassical character, meaning that the fluctuation (e.g., spin-wave) corrections from states outside the restricted basis are heavily quenched by the large anisotropy spin gaps. In turn, this shows that the RVB phase remains stable in an extended range of parameters, irrespective of the actual nature of the ground state of the KHAF or the presence of a gap. In particular, our finite-size results suggest that the RVB range can extend up to relatively large values of $|K|\!\sim\!J$, where $J$ is  the Heisenberg coupling.
 
The remaining part of the paper is organized as follows. We begin in Sec.~\ref{sec:Modeletc} with a general discussion of the model, its symmetries and duality transformations. In Sec.~\ref{sec:ClassicalPD} we revisit the classical phase diagram using the Luttinger-Tisza approach~\cite{LT1946,Bertaut1961,Litvin1974}. This approach leads naturally to a simple parametrization of the classical GS manifold in terms of Ising-like variables. 
In Sec.~\ref{sec:ObD} we present a semiclassical perturbative expansion that reveals the order by disorder mechanism operating in a large part of the phase diagram.  
In Sec.~\ref{sec:QuantumED} we present our extensive numerical study of the quantum spin $S\!=\!1/2$ model, obtained from exact diagonalizations in two different bases, one in the full basis and the other in the restricted orthonormalized basis of classical GSs. 
In Sec.~\ref{sec:Discussion} we provide a discussion and a broader perspective of our study. Technical details and auxiliary information are relegated to three Appendices (App.~\ref{app:LambdaMatrices}-\ref{app:ED24site}).

\vspace*{-0.3cm}
\section{Model, dualities and symmetries}\label{sec:Modeletc}
\vspace*{-0.3cm}
\subsection{Model}\label{sec:Model} 
\vspace*{-0.3cm}
We consider interacting spins ${\bf S}_i$ residing at the vertices $i$ of the 2D kagome lattice, a portion of which is shown in Fig.~\ref{fig:Model}. The kagome has a triangular Bravais lattice and a basis of three sites, A, B and C (shaded triangles). The nearest neighbour (NN) bonds of the lattice can be divided into three types, `xx' (red), `yy' (green) and `zz' (blue), depending on their orientation, see Fig.~\ref{fig:Model}. 
The Heisenberg-Kitaev or $JK$-model is described by the spin Hamiltonian
\be\label{eq:Model}
\mc{H} = \sum\nolimits_{\langle ij\rangle} \left( J ~\vec{S}_i\!\cdot\!\vec{S}_j + K ~S_i^{\gamma_{ij}}S_j^{\gamma_{ij}} \right)\,.
\ee
Here $\langle ij\rangle$ denotes NN lattice sites, ${\bf S}_i$ and ${\bf S}_j$ are the associated spin-1/2 degrees of freedom residing on these sites, and $J$ and $K$ denote the Heisenberg and Kitaev exchange couplings, respectively. The Cartesian components $\gamma_{ij}$ appearing in the Kitaev coupling equals $x$, $y$ or $z$,  depending on whether $\langle ij\rangle$ belongs to the `xx', `yy' or `zz' bond type. 
In the following, we measure energy in units of $J^2+K^2\!=\!1$ and parametrize
\be\label{eq:psi}
J=\cos\psi\,,~~ K=\sin\psi\,,~~\psi\in [0,2\pi)\,.
\ee

\vspace*{-0.3cm}
\subsection{Global symmetries}\label{sec:GlobalSymmetries}
\begin{figure}[!t]
\includegraphics[width=0.9\linewidth]{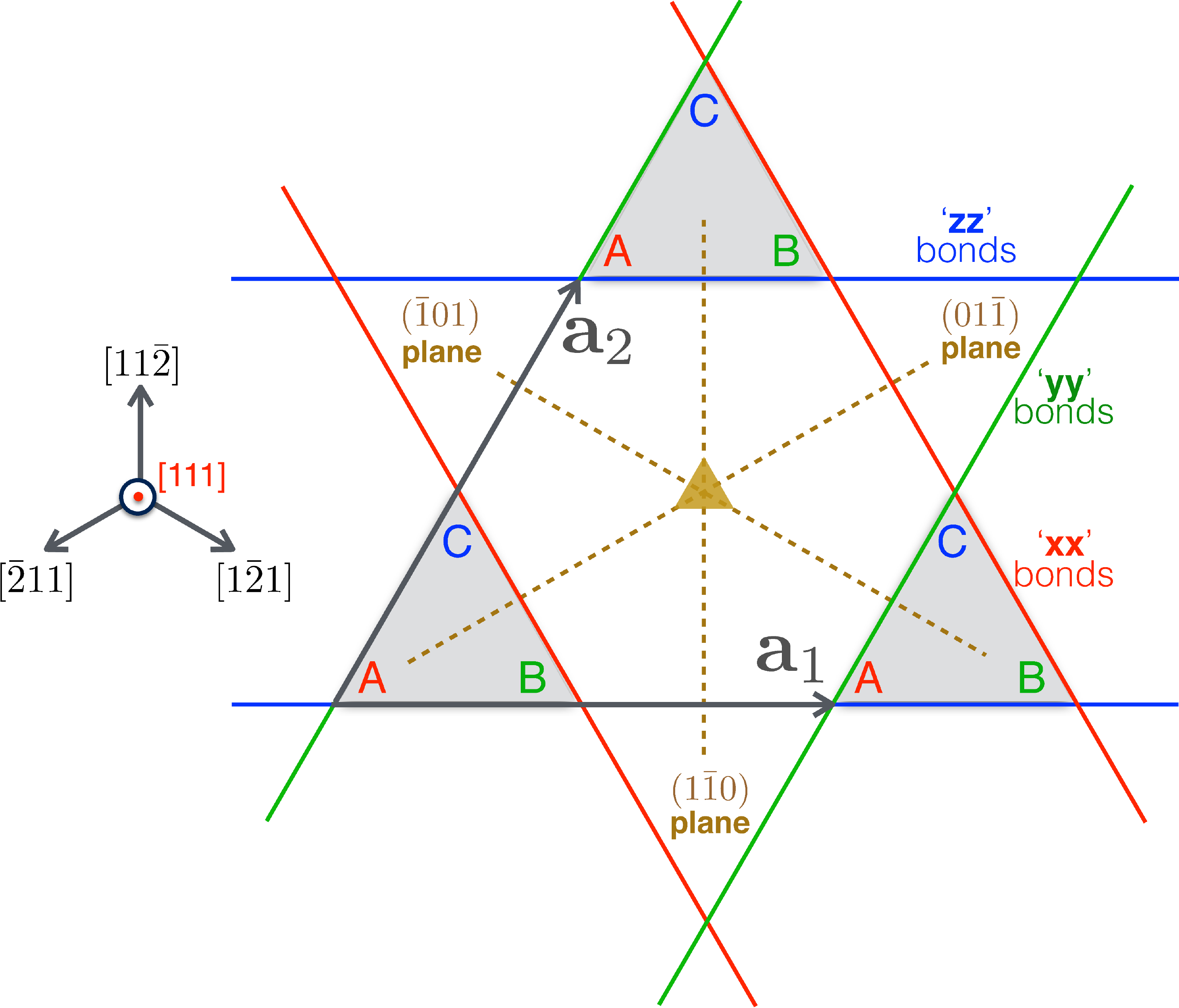}
\vspace*{-0.3cm} 
\caption{\label{fig:Model} The kagome lattice has a basis of three sites (A, B and C) and a triangular Bravais lattice with ${\bf a}_1$ and ${\bf a}_2$ denoting two primitive translations. Red, green and blue bonds between NN sites $(i,j)$ carry three distinct Kitaev interactions, $S^x_iS^x_j$, $S^y_iS^y_j$ and $S^z_iS^z_j$, respectively. With our choice of reference frame, the lattice sits on the $(111)$ plane and ${\bf a}_1\!=\!a\frac{{\bf x}-{\bf y}}{\sqrt{2}}$, ${\bf a}_2\!=\!a\frac{{\bf x}-{\bf z}}{\sqrt{2}}$, where $a$ is a lattice constant.}
\end{figure}

\vspace*{-0.3cm}
For half-integer spins, the Hamiltonian (\ref{eq:Model}) is in general invariant under the global symmetry group $\mc{T}\times \widetilde{C}_{3\text{v}} \times \widetilde{\mathsf{D}}_2$, which consists of the following operations: 

i) The translation group $\mc{T}$ generated by the primitive translation vectors $\vec{a}_1$ and $\vec{a}_2$ shown in Fig.~\ref{fig:Model}. 

ii) The double cover $\widetilde{\mathsf{C}}_{3\text{v}}$ of the group $\mathsf{C}_{3\text{v}}\!\subset\!\mathsf{SO}(3)$ in spin-orbit space, with the three-fold axis going through one of hexagon centers (see Fig.~\ref{fig:Model}). This is a three-fold rotation around $[111]$, which maps `xx', `yy' and `zz' bonds into `yy', `zz' and `xx' bonds in real space, and $(S^x,S^y,S^z)\mapsto(S^y,S^z,S^x)$  in spin space. The reflection planes $(1\bar{1}0)$, $(01\bar{1})$ and $(\bar{1}01)$, are shown by dashed (brown) lines in Fig.~\ref{fig:Model}. In spin space, a reflection through $(1\bar{1}0)$ maps $(S^x,S^y,S^z)\mapsto(-S^y,-S^x,-S^z)$, and similarly for the other planes. 

iii) The double cover $\widetilde{\mathsf{D}}_2$ of the point group $\mathsf{D}_2\subset\mathsf{SO}(3)$ in spin space alone. This consists of the three $\pi$-rotations $\mathsf{C}_{2x}$, $\mathsf{C}_{2y}$, and $\mathsf{C}_{2z}$, which map $(S^x,S^y,S^z)$ to $(S^x,-S^y,-S^z)$, $(-S^x,S^y,-S^z)$ and $(-S^x,-S^y,S^z)$, respectively.

The Hamiltonian has additional, hidden symmetries (self-dualities) at special points in parameter space, see discussion at the end of Sec.~\ref{sec:ThreeSubDualities} and Sec.~\ref{sec:KitaevSelfDualities}.

\vspace*{-0.3cm}
\subsection{Three-sublattice dualities}\label{sec:ThreeSubDualities}
\begin{figure*}[!t]
\includegraphics[width=\linewidth]{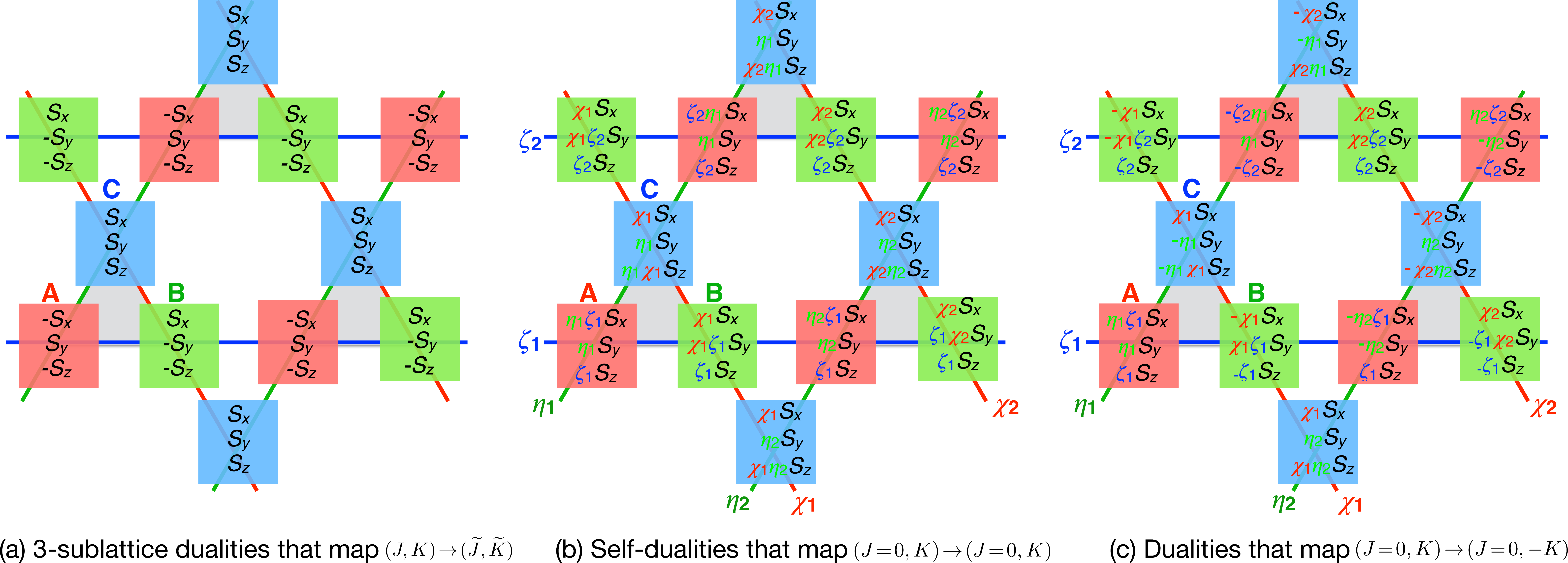}
\caption{\label{fig:Dualities} Transformed spin operators $\widetilde{{\bf S}}_i$ for each given kagome site $i$ under: (a) the 3-sublattice duality $\mathsf{D}_C$, (b) the sub-extensive self-dualities of the Kitaev points (Sec.~\ref{sec:KitaevSelfDualities}), and (c) the sub-extensive dualities that map one Kitaev point to the other (Sec.~\ref{sec:KitaevDualities}). 
In (b) and (c), each duality is characterized by a set of $3L$ Ising-like variables $\pm1$:  $\{\chi_1, \chi_2, \cdots, \chi_L\}$ for the `xx' lines, $\{\eta_1, \eta_2, \cdots, \eta_L \}$ for the `yy' lines, and $\{\zeta_1, \zeta_2, \cdots, \zeta_L\}$ for the `zz' lines, where $L$ is the number of lines of each type (the linear size of the system).} 
\end{figure*}

\vspace*{-0.3cm}
Similarly to the $JK$-model on other lattices,\cite{Jackeli2009,Chaloupka2010,Kimchi2014,Chaloupka2015,Rousochatzakis2015,Rousochatzakis2016,Becker2015,Chaloupka2019} the Hamiltonian (\ref{eq:Model}) supports duality transformations -- also referred to as the Klein duality~\cite{Kimchi2014} -- which preserve the form of the Hamiltonian but alter the value of the parameters $J$ and $K$. 
As shown in Ref.~\cite{Kimchi2014}, the main difference with the other lattices is that here the duality transformations involve three sublattices A, B and C (see Fig.\ref{fig:Model}) instead of four. 
There are three such transformations, which we denote by $\mathsf{D}_{A}$, $\mathsf{D}_{B}$ and $\mathsf{D}_{C}$, and consist of a combination of $\pi$-rotations around the $x$-, $y$- or $z$-axis, depending on the sublattice index. For example, $\mathsf{D}_C$ arises by a product of $\mathsf{C}_{2y}$ rotations for the A sites and $\mathsf{C}_{2x}$ rotations for the B sites,
\be\label{eq:DualityDC}
\begin{array}{l}
\mathsf{D}_{\text{C}}=\prod_{i \in A} \mathsf{C}_{2y}(i)\cdot\prod_{j \in B} \mathsf{C}_{2x}(j)\,.
\end{array}
\ee
Under $\mathsf{D}_C$ the various spin operators ${\bf S}_i$ transform to 
\be\label{equ:duality}
\widetilde{{\bf S}}_i=\left\{
\renewcommand{\arraystretch}{1.25}
\begin{array}{ll}
( -S^x_i,\, S^y_i,\, -S^z_i ),	& \text{if}~~i\in~\text{A}\\
( S^x_i,\, -S^y_i,\, -S^z_i ),	& \text{if}~~i\in~\text{B}\\
( S^x_i,\, S^y_i,\, S^z_i ),	& \text{if}~~i\in~\text{C}
\end{array}\,,
\right.
\ee
as shown in Fig.~\ref{fig:Dualities}\,(a). 
Similarly, $\mathsf{D}_A$ and $\mathsf{D}_B$ are given by 
\be
\renewcommand{\arraystretch}{1.25}
\begin{array}{l}
\mathsf{D}_{A}\!=\!\prod_{i \in B}\!\mathsf{C}_{2z}(i)\!\cdot\!\prod_{j \in C}\!\mathsf{C}_{2y}(j),\\
\mathsf{D}_{B}\!=\!\prod_{i \in A}\!\mathsf{C}_{2z}(i)\!\cdot\!\prod_{j \in C}\!\mathsf{C}_{2x}(j).
\end{array}
\ee
Under $\mathsf{D}_A$, $\mathsf{D}_B$ or $\mathsf{D}_C$, the Hamiltonian maps to 
\begin{align}\label{eq:Modeldual}
\widetilde{\mc{H}} = \sum\nolimits_{\langle ij\rangle} \left( -J ~\widetilde{\vec{S}}_i\!\cdot\!\widetilde{\vec{S}}_j + (2J+K) ~\widetilde{S}_i^{\gamma_{ij}}\widetilde{S}_j^{\gamma_{ij}} \right)\,,
\end{align}
i.e., the general form of (\ref{eq:Model}) is preserved but $K$ and $J$ map to
\be\label{eq:KtildeJtilde}
\widetilde{K}=2J+K ~~\text{and}~~ \widetilde{J}=-J\,.
\ee 
The duality transformations allow to identify special symmetry points of the parameter space and relate different regions of the phase diagram of Fig.~\ref{fig:PhaseDiagram} (which will be discussed in detail in Sec.~\ref{sec:ClassicalPD} below). 
Specifically, the region `IA' of Fig.~\ref{fig:PhaseDiagram} maps to the region `IB', while the region `IIA' maps to `IIB'. 
Moreover, the Heisenberg points $\psi\!=\!0$ and $\pi$, where $K\!=\!0$, map to the dual Heisenberg points $\psi\!=\!\pi-\arctan{2}\!\simeq\!0.6475\pi$ and $-\arctan{2}\!\simeq\!1.6475\pi$, respectively, where $\widetilde{K}\!=\!0$. These dual points have therefore a hidden SU(2) rotation symmetry. 
The Kitaev points $\psi\!=\!\pm\pi/2$ are self-dual points, since  at these points $\widetilde{J}\!=\!J\!=\!0$, $\widetilde{K}\!=\!K$ and $\widetilde{\mc{H}}\!=\!\mc{H}$, i.e., the transformations become symmetries.

\begin{figure}[!t]
\includegraphics[width=0.95\linewidth]{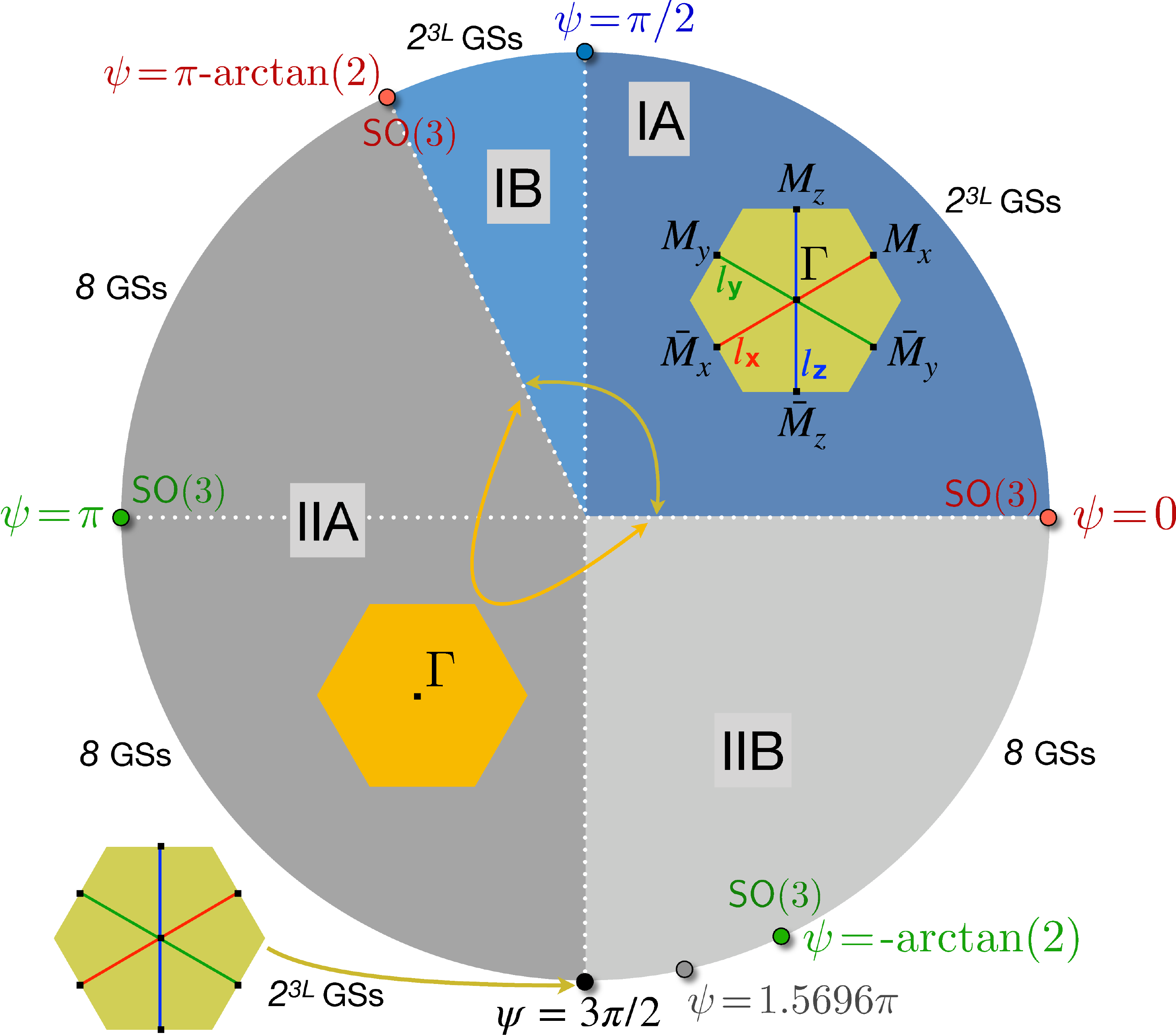}
\caption{\label{fig:PhaseDiagram} Classical phase diagram of the $JK$-model (\ref{eq:Model}) parametrized by the angle $\psi$ of Eq.~(\ref{eq:psi}). The inset hexagons show the ${\bf k}$-points of the first Brillouin zone associated with the minimum eigenvalue of the coupling matrix $\bs{\Lambda}_{\bf k}$.} 
\end{figure}

\vspace*{-0.3cm}
\subsection{Self-dualities at the Kitaev points}\label{sec:KitaevSelfDualities} 
\vspace*{-0.3cm}
As it turns out, the two Kitaev points have many more self-dualities (than $\mathsf{D}_A$, $\mathsf{D}_B$, $\mathsf{D}_C$), in fact, a subextensive number of them. The presence of these symmetries has not been recognized in previous studies, and explains naturally the absence of classical and quantum order by disorder at these special points~\cite{Morita2018,Morita2019b}.

More specifically, there are in total $2^{3L}$ self-duality transformations that map $(J=0, K)\mapsto (J=0, K)$, where $3L$ is the total number of lines in the lattice.  
The general form of the transformed spin operators under these operations is shown in Fig.~\ref{fig:Dualities}\,(b). Each transformation is characterized by a set of $3L$ Ising-like variables $\pm 1$, one for each line of the lattice. In Fig.~\ref{fig:Dualities}\,(b), these variables are denoted by $\{\chi_1,\chi_2,\cdots,\chi_L\}$ for `xx' lines (red),  $\{\eta_1,\eta_2,\cdots,\eta_L\}$ for `yy' lines (green), and  $\{\zeta_1,\zeta_2,\cdots,\zeta_L\}$ for `zz' lines (blue). 
Note that changing the sign of any of these variables corresponds to flipping two spin components for every site residing on the corresponding line. 
For the horizontal line of Fig.~\ref{fig:Dualities}\,(b) associated with the number $\zeta_1$, for example, changing $\zeta_1\!\mapsto\!-\zeta_1$ amounts to a product of alternating $\mathsf{C}_{2y}$ and $\mathsf{C}_{2x}$ rotations along this line. One can check that the Hamiltonian remains invariant under this operation. 

Similar subextensive symmetries appear in the {\it quantum} compass model on the square~\cite{Doucot2005,Dorier2005}, cubic~\cite{Nussinov2015} and honeycomb~\cite{Chen2008} lattices, although the analogous operations of flipping individual line variables amount to products of $\mathsf{C}_{2}$ rotations around a {\it fixed} axis (and not around alternating axes as here). 
One should be cautious to differentiate here from the case of subextensive operations appearing in certain {\it classical} models but are absent from their {\it quantum} counterparts. This includes, for example, the Kitaev model on the triangular lattice~\cite{Rousochatzakis2016} and the $K_1$-$K_2$ model on the honeycomb lattice~\cite{Rousochatzakis2015}. In these cases, the subextensive operations involve flipping one Cartesian spin component, which is a valid operation only for classical spins~\footnote{The reason being that to flip the sign of a single component we must combine a $\pi$-rotation with time reversal, but the latter acts on the whole system and not on the sites of the line alone~\cite{Rousochatzakis2015,Rousochatzakis2016}.}, whereas here the transformed spin components shown in Fig.~\ref{fig:Dualities}\,(b) always correspond to flipping two components.

Now, the important point about the above subextensive quantum symmetries is that they affect only a line of spin sites, and are therefore intermediate between global operations and local (single-site) operations. A generalization of Elitzur's theorem~\cite{Elitzur1975} by Batista and Nussinov~\cite{Batista2005} then asserts that in 2D such symmetries can be broken spontaneously only at zero temperature. This implies that the kagome Kitaev model can host a zero-temperature long-range ordered phase, and the latter can be anyone among $2^{3L}$ different degenerate GSs. 
Diagnosing this order in finite-size calculations, however, is far from straightforward, as we explain in detail in Sec.~\ref{sec:Correlations}.

We note finally that the 3-sublattice dualities $\mathsf{D}_A$, $\mathsf{D}_B$ and $\mathsf{D}_C$ are special members of the subextensive family of self-dualities. Specifically, they correspond to the following choice of $\chi_\ell$, $\eta_\ell$ and $\zeta_\ell$ ($\ell\!=\!1\cdots L$):
\be
\renewcommand{\arraystretch}{1.25}
\begin{array}{l}
\mathsf{D}_A:~~ (\chi_\ell, \eta_\ell,\zeta_\ell)=(-1,1,1)\,,~~~ \forall \ell=1\cdots L\,, \\
\mathsf{D}_B:~~ (\chi_\ell, \eta_\ell,\zeta_\ell)=(1,-1,1)\,,~~~ \forall \ell=1\cdots L\,, \\
\mathsf{D}_C:~~ (\chi_\ell, \eta_\ell,\zeta_\ell)=(1,1,-1)\,,~~~ \forall \ell=1\cdots L\,.
\end{array}
\ee

\vspace*{-0.3cm}
\subsection{Dualities that map $(J\!=\!0,K)\mapsto(J\!=\!0,-K)$}\label{sec:KitaevDualities}
\vspace*{-0.3cm}
A simple modification of the above self-dualities gives rise to a sub-extensive family of transformations that map one Kitaev point to another, namely $(J\!=\!0,K)\mapsto(J\!=\!0,-K)$. The corresponding form of the transformed spin operators in this family of dualities is shown in Fig.~\ref{fig:Dualities}\,(c). 
Compared to the self-dualities shown in Fig.~\ref{fig:Dualities}\,(b), here the signs of the $3L$ numbers $\chi_\ell$, $\eta_\ell$ and $\zeta_\ell$ ($\ell\!=\!1\cdots L$) alternate between $+1$ and $-1$ along their corresponding lines.
Doing this for all lines results in effectively reversing the sign of the Kitaev coupling for all NN bonds. 
We learn therefore that the physics of the AF Kitaev model can be mapped to the physics of the ferromagnetic (FM) Kitaev model.

\vspace*{-0.3cm}
\section{Classical phase diagram}\label{sec:ClassicalPD}
\vspace*{-0.3cm}
Some aspects of the classical phase diagram have been discussed by Kimchi and Vishwanath~\cite{Kimchi2014}, and a more complete characterization has been given more recently by Morita, Kishimoto and Tohyama~\cite{Morita2018} starting by the classical minimum of a triangular unit cell and tiling the solution to the lattice. We shall give here a complementary picture based on the Luttinger-Tisza (LT) approach~\cite{LT1946,Bertaut1961,Litvin1974,Kaplan2007}, which leads naturally to a parametrization of the GSs that will prove convenient for the discussion of the order by disorder effect.

\vspace*{-0.3cm}
\subsection{General setting of LT approach}\label{sec:LTgeneral}
\vspace*{-0.3cm}
In the LT approach one replaces the problem of minimizing the energy $E$ under $\mc{N}$ `hard' spin length constraints (${\bf S}_i^2\!=\!S^2$, $i\!=\!1$-$\mc{N}$, where $\mc{N}$ is the total number of sites) by the much simpler problem of minimizing $E$ under a single `soft' constraint $\sum_i {\bf S}_i^2\!=\!\mc{N}S^2$. The latter problem then reduces to the diagonalization of a coupling matrix $\bs{\Lambda}$, whose lowest eigenvalue corresponds to the lowest energy of the soft problem. It then follows~\cite{LT1946,Bertaut1961,Litvin1974,Kaplan2007} that if an eigenvector corresponding to the minimum eigenvalue can be used to construct a configuration that satisfies the $\mc{N}$ original constraints then that configuration must be one of the GSs of the hard problem. As we will now show, this approach works successfully in the present model. 

We begin by labelling the spin sites by $({\bf R},\mu)$, where ${\bf R}\!=\! n_1{\bf a}_1+n_2{\bf a}_2$ (where $n_1$ and $n_2$ are integers) denotes the positions of the unit cell, and $\mu\!\in\!\{$A, B, C$\}$ is the sublattice index. We next go to momentum space and define 
\be
{\bf S}_{\vec{k},\mu}\!=\!\frac{1}{\mc{N}/3}\sum_{\bf R} e^{-i \vec{k}\cdot\vec{R}}\vec{S}_{\vec{R},\mu}\,,~~~
{\bf S}_{\vec{R},\mu}\!=\!\sum_{\bf k}  e^{i \vec{k}\cdot\vec{R}}\vec{S}_{\vec{k},\mu}\,,
\ee
where ${\bf k}$ belongs to the first Brillouin zone of the triangular Bravais lattice. We then rewrite $E/\mc{N}$ in a matrix form 
\be\label{eq:energy1}
E/\mc{N}=\frac{1}{2}\sum_{\vec{k}\in\text{BZ}} \mc{S}_{\bf k}^\dagger  \cdot \bs{\Lambda}_\vec{k} \cdot \mc{S}_{\bf k}\,,
\ee
where $\mc{S}_{\bf k}$ is the $9\times1$ vector 
\be
\!\!\!\!\mc{S}_{\bf k}\!=\!\left(S^x_{{\bf k},A}, S^y_{{\bf k},A}, S^z_{{\bf k},A}, S^x_{{\bf k},B}, S^y_{{\bf k},B}, S^z_{{\bf k},B}, S^x_{{\bf k},C}, S^y_{{\bf k},C}, S^z_{{\bf k},C}\right)^T\!,
\ee
where the $9\times9$ coupling matrix $\bs{\Lambda}_{\bf k}$ has a block-diagonal form
$\bs{\Lambda}_{\bf k}=\text{diag}(\Lambda^{(x)}_{\bf k},\Lambda^{(y)}_{\bf k},\Lambda^{(z)}_{\bf k})$, and the $3\times3$ matrices $\Lambda^{(\alpha)}_{\bf k}$ ($\alpha\!=\!x$, $y$, $z$) are provide in App.~\ref{app:LambdaMatrices}.
The coupling matrix is hermitian and has therefore a complete set of orthonormal eigenvectors $V_{{\bf k},\nu}$ (with $\nu=1$-$9$) satisfying
\be
\bs{\Lambda}_{\bf k} V_{{\bf k},\nu} = \lambda_{{\bf k},\nu} V_{{\bf k},\nu}\,.
\ee
We can then decompose $\bs{\Lambda}_{\bf k}=\sum_\nu \lambda_{{\bf k},\nu} V_{{\bf k},\nu} V^\dagger_{{\bf k},\nu}$ and rewrite
\be\label{eq:energy2}
E/\mc{N} = \frac{1}{2}\sum_{{\bf k},\nu} \lambda_{{\bf k},\nu} |c_{{\bf k},\nu}|^2\,,
\ee
where the coefficients $c_{{\bf k},\nu}$ are defined via
\be
\mc{S}_{\bf k}=\sum_\nu c_{{\bf k},\nu} V_{{\bf k},\nu},~~
c_{{\bf k},\nu}=V^\dagger_{{\bf k},\nu}\cdot \mc{S}_{{\bf k}}\,.
\ee
Equation~(\ref{eq:energy2}) then tells us that we can saturate the minimum of the energy by using the modes at the special points $({\bf k}^\ast,\nu^\ast)$ associated with the minimum eigenvalue $\lambda_{\text{min}}$, i.e., to replace 
\be
\mc{S}_{\bf R}\!\equiv\!\sum_{\bf k}e^{i{\bf k}\cdot{\bf R}}\mc{S}_{\bf k} ~~\to~~ 
\sum_{{\bf k}^\ast, \nu^\ast}\! c_{{\bf k}^\ast,\nu^\ast} e^{i {\bf k}^\ast\cdot{\bf R}} V_{{\bf k}^\ast,\nu^\ast}\,.
\ee
The energy then becomes 
\be
\label{eq:energy3}
E/\mc{N} = \frac{1}{2}\lambda_{\text{min}} \sum_{{\bf k}^\ast,\nu^\ast} |c_{{\bf k}^\ast,\nu^\ast}|^2 = \frac{3S^2}{2}\lambda_{\text{min}}\,,
\ee
where in the second step we used the soft constraint in momentum space, $\sum_{{\bf k}^\ast,\nu^\ast}\!|c_{{\bf k}^\ast,\nu^\ast}|^2\!=\!3S^2$. 
As we show next, the coefficients $c_{{\bf k}^\ast,\nu^\ast}$ can be chosen in such a way as to satisfy the spin length constraints.

\vspace*{-0.3cm}
\subsection{Classical ground states inside the regions IA and IB}\label{sec:ClassGSsIAandIB}
\begin{figure}[!t]
\centering
\includegraphics[width=\linewidth]{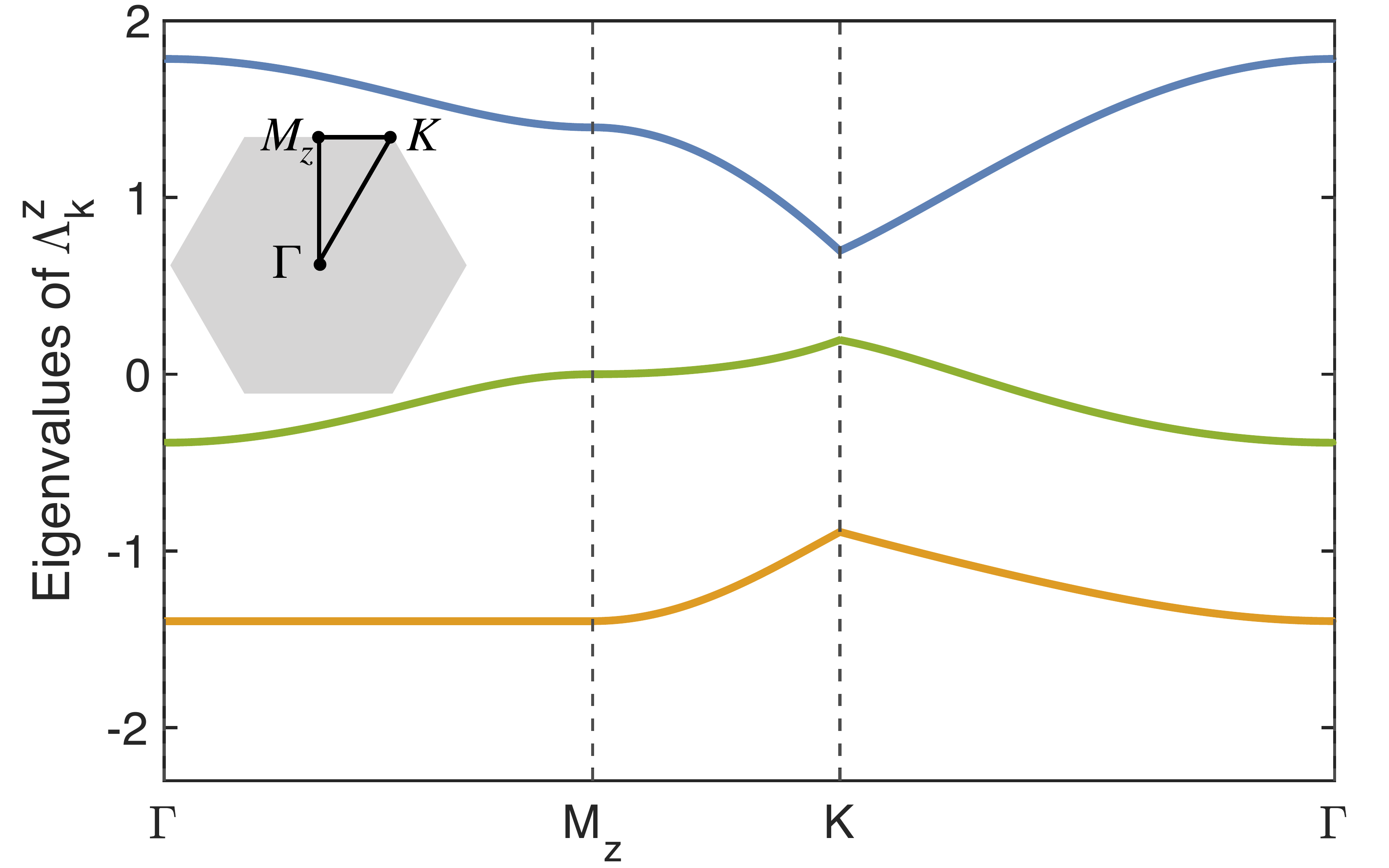}
\caption{Evolution of the three eigenvalues of $\Lambda^{(z)}_{\bf k}$ at $\psi\!=\!0.3\pi$ (region IA) along the first Brillouin zone symmetry paths shown in the inset.}\label{fig:phase1Luttinger}
\end{figure}

\vspace*{-0.3cm}
Inside the regions IA and IB the minimum eigenvalue is achieved along the three special lines that connect opposite $M$ points of the Brillouin zone, see representative case shown in Fig.~\ref{fig:phase1Luttinger}. These lines are denoted by $l_x$, $l_y$ and $l_z$ in the upper inset (yellow) hexagon of Fig.~\ref{fig:PhaseDiagram}. The corresponding eigenvectors take the following simple form
\be
\renewcommand{\arraystretch}{1.25}
\begin{array}{l}
{\bf k}^\ast\!\in\! l_x, \nu^\ast\!=\!1: ~~ V_{l_x,1} =\frac{1}{\sqrt{2}}(0,0,0,-1,0,0,1,0,0)^T\,,\\
{\bf k}^\ast\!\in\! l_y, \nu^\ast\!=\!4: ~~ V_{l_y,4}=\frac{1}{\sqrt{2}}(0,-1,0,0,0,0,0,1,0)^T\,,\\
{\bf k}^\ast\!\in\! l_z, \nu^\ast\!=\!7: ~~ V_{l_z,7}=\frac{1}{\sqrt{2}}(0,0,-1,0,0,1,0,0,0)^T\,.
\end{array}
\ee
We can then build GSs by linearly combining these eigenvectors, leading to  
\be\label{eq:phaseI}
\begin{array}{c}
{\vec S}_{\mathbf{R},\text{A}}\!=\!
\frac{S}{\sqrt{2}}\left(\!\!
\begin{array}{c}
0 \\
-\eta_{\bf R}\\
-\zeta_{\bf R}
\end{array}
\!\!\right),~
{\vec S}_{\mathbf{R},\text{B}}\!=\!
\frac{S}{\sqrt{2}}\left(\!\!
\begin{array}{c}
-\chi_{\bf R}\\
0\\
\zeta_{\bf R}
\end{array}
\!\!\right),~
{\vec S}_{\mathbf{R},\text{C}}\!=\!
\frac{S}{\sqrt{2}}\left(\!\!
\begin{array}{c}
\chi_{\bf R}\\
\eta_{\bf R} \\
0
\end{array}
\!\!\right)
\end{array},
\ee
where 
\be
\chi_{\bf R}\!\equiv\!\!\sum_{\mathbf{k}\in l_x}\!c_{{\bf k},1}e^{i {\bf k}\cdot{\bf R}},~~
\eta_{\bf R}\!\equiv\!\!\sum_{\mathbf{k}\in l_y}\!c_{{\bf k},4}e^{i {\bf k}\cdot{\bf R}},~~ 
\zeta_{\bf R}\!\equiv\!\!\sum_{\mathbf{k}\in l_z}\!c_{{\bf k},7}e^{i {\bf k}\cdot{\bf R}}.
\ee
Imposing ${\bf S}_{{\bf R},\text{A}}^2\!=\!{\bf S}_{{\bf R},\text{B}}^2\!=\!{\bf S}_{{\bf R},\text{C}}^2\!=\!S^2$ gives 
\be
\chi_{\bf R}^2\!=\!\eta_{\bf R}^2\!=\!\zeta_{\bf R}^2\!=\!1\,,
\ee 
i.e., the coefficients $\chi_{\bf R}$, $\eta_{\bf R}$ and $\zeta_{\bf R}$ are constrained to $+1$ or $-1$. 
Note further that from the extensive set of different choices of $\chi_{\bf R}$ as we vary ${\bf R}$ (and similarly for $\eta_{\bf R}$ and $\zeta_{\bf R}$), only a sub-extensive subset are independent. This stems from the fact that 
\be
\text{for}~~\bs{\delta}\parallel \text{`xx' lines}:~~~
\chi_{{\bf R}+\bs{\delta}}\!=\!
\sum_{\mathbf{k}\in l_x}\!c_{{\bf k},1}e^{i {\bf k}\cdot({\bf R}+\bs{\delta})}\!=\!\chi_{\bf R}\,,
\ee 
since the line $l_x$ of the Brillouin zone is vertical to the direction of the `xx' lines (see Figs.~\ref{fig:Model} and \ref{fig:PhaseDiagram}).  
Similarly, $\eta_{{\bf R}+\bs{\delta}}\!=\!\eta_{\bf R}$ for $\bs{\delta}\!\parallel$`yy' lines, and $\zeta_{{\bf R}+\bs{\delta}}\!=\!\zeta_{\bf R}$ for $\bs{\delta}\!\parallel$`zz' lines. 
Hence, for any given GS configuration, the coefficients $\chi_{\bf R}$, $\eta_{\bf R}$ and $\zeta_{\bf R}$ are fixed along individual `xx', `yy' or `zz' lines, respectively.

\begin{figure*}[!t]
\includegraphics[width=0.9\linewidth]{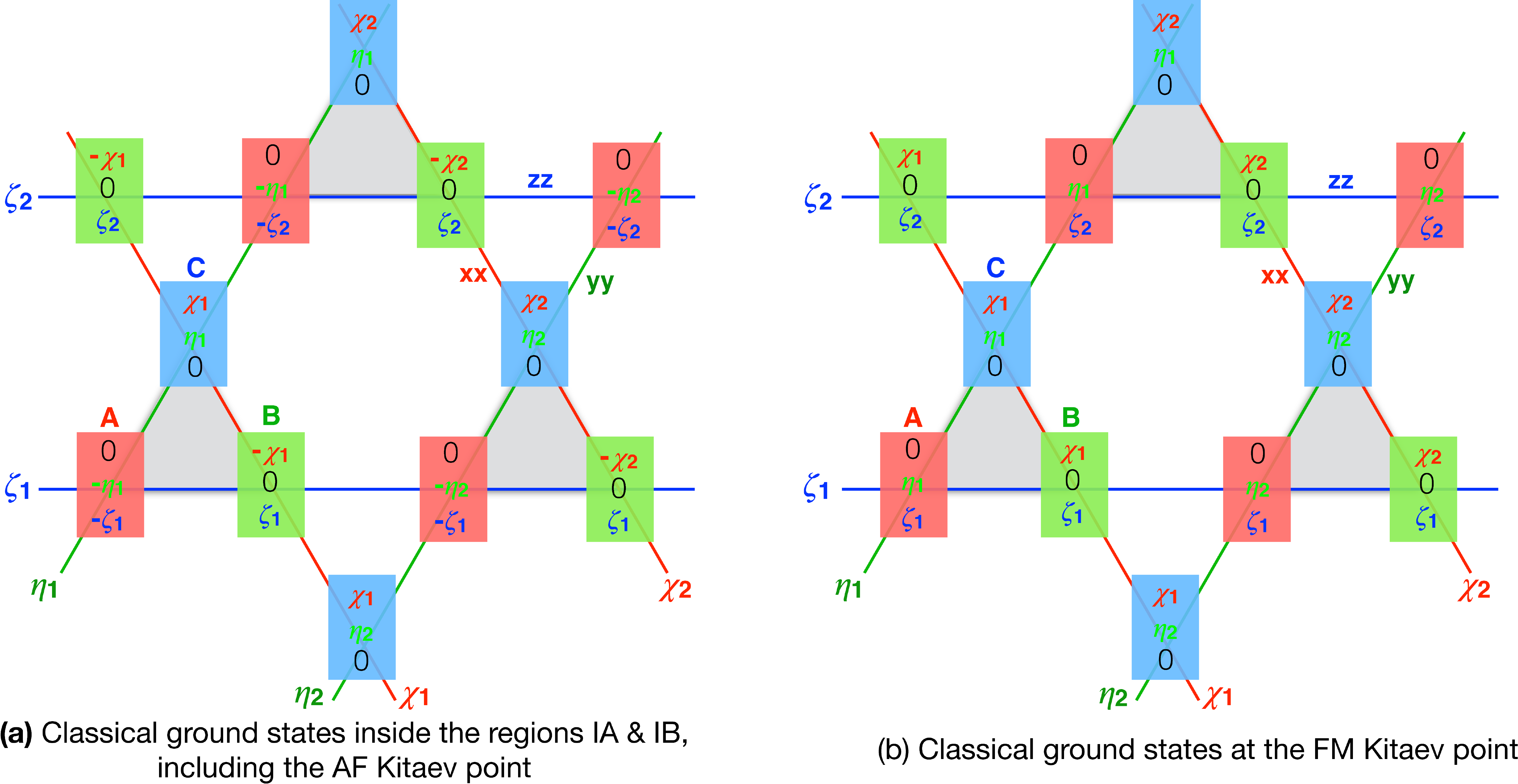}
\caption{\label{fig:IAIBandFMKitaev} General form of classical GSs inside the regions IA and IB of Fig.~\ref{fig:PhaseDiagram}\,(a) and at the FM Kitaev point (b). The numbers $\chi_\ell$, $\eta_\ell$ and $\zeta_\ell$ (where $\ell\!=\!1\cdots L$) are constrained to $+1$ or $-1$. The overall normalization prefactors of $S/\sqrt{2}$ have been omitted.} 
\end{figure*}

The general structure of the resulting states is shown in Fig.~\ref{fig:IAIBandFMKitaev}\,(a). Each state is characterized by a set of $3L$ Ising-like variables $\chi_\ell$, $\eta_\ell$ and $\zeta_\ell$  ($\ell\!=\!1\cdots L$).
Flipping the sign of one of these coefficients amounts to flipping the associated component for all sites on the corresponding line. Clearly, the total number of states is $2^{3L}$. This type of sub-extensive degeneracy is accidental, except for the AF Kitaev point where the degeneracy arises from the sub-extensive self-dualities discussed in Sec.~\ref{sec:KitaevSelfDualities} and shown in Fig.~\ref{fig:Dualities}\,(b). 

A few comments on the general structure of the GSs are in order here. First, the total spin in each unit cell of the lattice vanishes, i.e., 
\be
\text{region IA \& IB}:~~~ 
{\bf S}_{{\bf R},\text{A}}+{\bf S}_{{\bf R},\text{B}}+{\bf S}_{{\bf R},\text{C}}=0\,,
\ee
as can be seen from Eq.~(\ref{eq:phaseI}). This is precisely the condition that minimizes the classical energy of the KHAF~\cite{HarrisKallinBerlinsky1992,Sachdev1992,Chubukov1993}. Therefore, the subextensive set of classical GSs of the regions IA and IB (including the AF Kitaev point) is a subset of the extensive classical GSs of the KHAF.
The three spins in each given unit cell are coplanar (with $(\chi_{\bf R}, -\eta_{\bf R}, \zeta_{\bf R})$ being the spin plane), and form an angle of 120$^\circ$ relative to each other. However the GSs are globally non-coplanar, in general. 

Second, the GSs remain non-collinear even at the AF Kitaev point. This is qualitatively different from what happens in the square compass~\cite{Dorier2005,Nussinov2015} and triangular Kitaev model~\cite{Rousochatzakis2016,Becker2015}, where spins align along one of the three Cartesian axes leading to non-zero spin correlations along the corresponding type of lines and vanishing correlations between different lines. Here such a decoupling between lines does not occur because of the special corner-sharing-triangle topology of the kagome.

Third, the spin components along a given Cartesian axis are correlated antiferromagnetically along the corresponding `xx', `yy' or `zz' line, see for example the alternation of $z$-components along the line associated with $\zeta_1$ in Fig.~\ref{fig:IAIBandFMKitaev}\,(a).

Finally, the expression for the classical GS energy inside the regions IA and IB is 
\be\label{eq:EminIA&IB}
\text{region IA \& IB}:~~~ 
E_{\text{min}}/\mc{N} = -(K+J) S^2\,,
\ee
which can be easily veriified by looking at the general structure of the GSs in Fig.~\ref{fig:IAIBandFMKitaev}\,(a) and by noting that each bond contributes an energy of $-(K+J)S^2/2$.

\vspace*{-0.3cm}
\subsection{Classical ground states inside the regions IIA and IIB}\label{sec:ClassGSsIIAandIIB}
\vspace*{-0.3cm}
Let us now turn to the GSs inside the regions IIA and IIB of Fig.~\ref{fig:PhaseDiagram}. We will first discuss what happens away from the special points $\psi\!=\!0.6475\pi$, $3\pi/2$ and $2\pi$. 
Here the minima of the eigenvalue spectrum of $\bs{\Lambda}_{\bf k}$ sits at the $\bs{\Gamma}$ point ${\bf k}\!=\!0$ of the Brillouin zone, with energy per site 
\be
\text{region IIA-IIB:}~~
E/\mc{N}\!=\!S^2 [ J\!+\!K\!-\!\sqrt{8J^2\!+\!(J\!+\!K)^2}~]/2\,,
\ee
The minimum eigenvalue is 3-fold degenerate, with corresponding eigenvectors:
\be\label{PhaseIIvectors}
\renewcommand{\arraystretch}{1.25}
\begin{array}{l}
{\bf k}^\ast\!=\!0, \nu^\ast\!=\!1: ~~ V_{\bs{\Gamma},1}=\xi (-\lambda,0,0,1,0,0,1,0,0)^T,\\
{\bf k}^\ast\!=\!0, \nu^\ast\!=\!2: ~~ V_{\bs{\Gamma},2}=\xi (0,1,0,0,-\lambda,0,0,1,0)^T,\\
{\bf k}^\ast\!=\!0, \nu^\ast\!=\!3: ~~ V_{\bs{\Gamma},3}=\xi (0,0,1,0,0,1,0,0,-\lambda)^T,
\end{array}
\ee
where $\xi\!=\!1/\sqrt{2\!+\!\lambda^2}$ and
\be\label{eq:lambda1}
\lambda\!=\!\left(\cos\psi\!+\!\sin\psi\!+\!\sqrt{5\!+\!4\cos 2\psi\!+\!\sin 2\psi}\right)/\left(2\cos\psi\right)\,.
\ee
A linear combination of these modes with coefficients $\chi_1$, $\eta_1$ and $\zeta_1$ (or $c_{\bs{\Gamma},1}$, $c_{\bs{\Gamma},2}$ and $c_{\bs{\Gamma},3}$ in the above notation) delivers the following structure for the GSs:
\small
\be\label{eq:ClassGSsIIAandIIB}
\!\!\!
{\vec S}_{{\bf R},\text{A}} = \xi S 
\left(\!\!\begin{array}{c}
-\lambda \chi_1 \\ \eta_1 \\ \zeta_1
\end{array}\!\!\right),~~ 
{\vec S}_{{\bf R},\text{B}} =  \xi S 
\left(\!\!\begin{array}{c}
\chi_1 \\ -\lambda \eta_1 \\ \zeta_1
\end{array}\!\!\right),~~
{\vec S}_{{\bf R},\text{C}} =  \xi S 
\left(\!\!\begin{array}{c}
\chi_1 \\ \eta_1 \\ -\lambda \zeta_1 
\end{array}\!\!\right)\,.
\ee
\normalsize
These expressions agree with the results of Morita {\it et al}~\cite{Morita2018} (where $\lambda$ is denoted by $f$).

\begin{figure}[!b]
\includegraphics[width=0.99\linewidth]{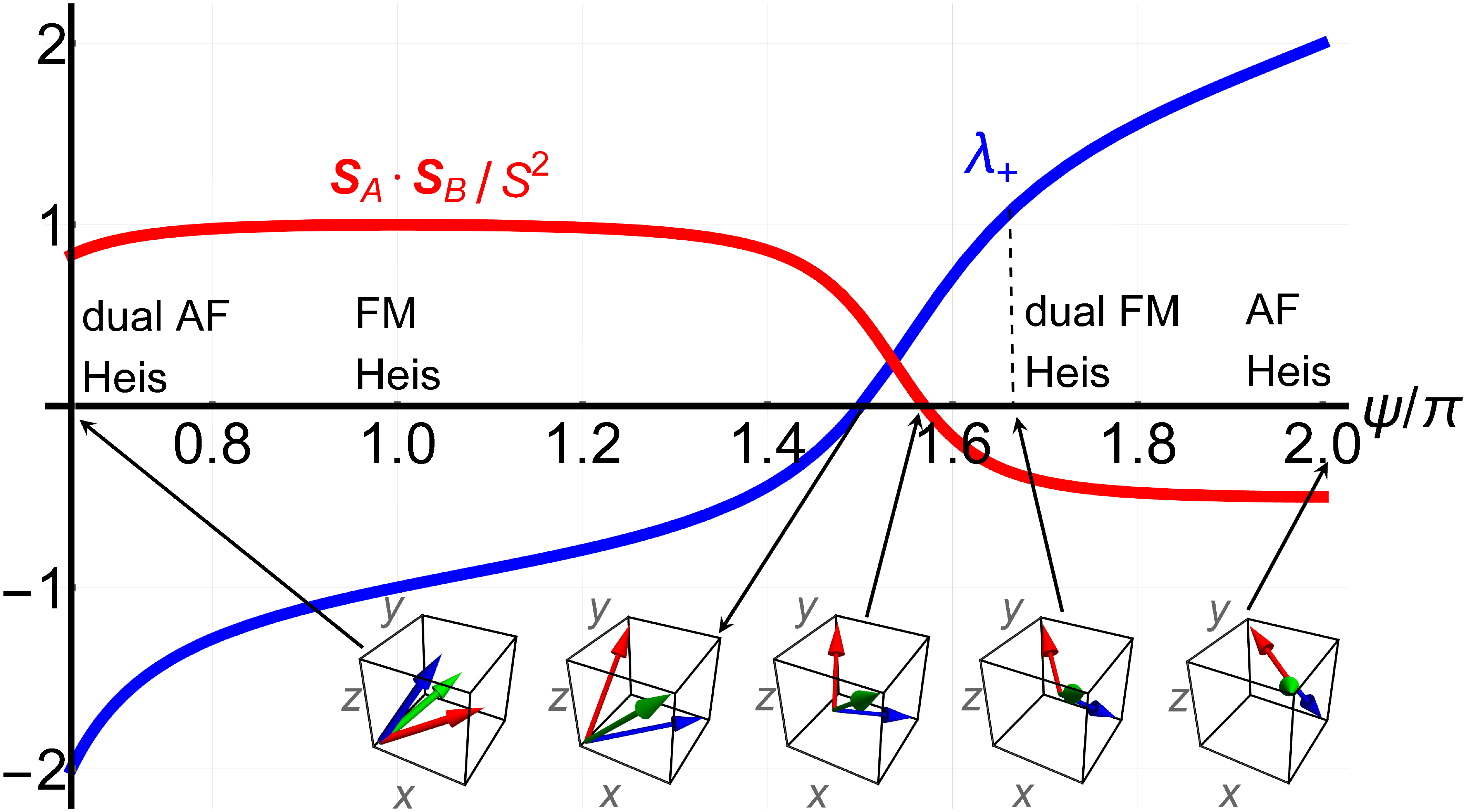}
\caption{Evolution of $\lambda$ [Eq.~(\ref{eq:lambda1})] and ${\bf S}_{\text{A}}\!\cdot\!{\bf S}_{\text{B}}$ [Eq.~(\ref{eq:SASB})]  with $\psi$ in the regions IIA and IIB of Fig.~\ref{fig:PhaseDiagram}. The insets show the three magnetic sublattices A  (red), B (green) and C (blue) for $\chi_1\!=\!\eta_1\!=\!\zeta_1\!=\!1$ and five representative values of $\psi$ ($0.6475\pi$, $1.5\pi$, $1.5696\pi$, $1.6475\pi$ and $2\pi$).}
\label{fig:SmuSnuAndLambdavspsi}
\end{figure}

Note that, unlike the regions IA and IB, here the GSs evolve with $\psi$, because $\lambda$ changes with $\psi$, see Fig.~\ref{fig:SmuSnuAndLambdavspsi}\,(blue curve).
Note also that the states are uniform (they do not depend on ${\bf R}$), they comprise three magnetic sublattices (A, B and C) and the angle between NN spins is common for all bonds, 
\be\label{eq:SASB}
\mathbf{S}_{\text{A}}\cdot\mathbf{S}_{\text{B}}
=\mathbf{S}_{\text{A}}\cdot\mathbf{S}_{\text{C}}
=\mathbf{S}_{\text{B}}\cdot\mathbf{S}_{\text{C}}
=S^2 (1-2\lambda)/(2+\lambda^2)\,.
\ee
The evolution of these correlations with $\psi$ is shown in Fig.~\ref{fig:SmuSnuAndLambdavspsi}\,(red curve). They are large and positive (i.e., FM) in most of the region IIA, they turn AF at $\tan\frac{\psi}{2}\!=\!\frac{2-\sqrt{85}}{9}$ ($\psi\!\simeq\!1.5695\pi$), and eventually approach the value $-1/2$ (i.e., the angle between NN spins tends to $120^\circ$) as $\psi\!\to\!2\pi$.

Let us next discuss the choice of the coefficients $\chi_1$, $\eta_1$ and $\zeta_1$. Imposing the spin length constraint on A, B and C spins gives the conditions
\be
(1-\lambda^2)\chi_1^2=(1-\lambda^2)\eta_1^2=(1-\lambda^2)\zeta_1^2\,,
\ee
which in turn lead to
\be
\left\{
\renewcommand{\arraystretch}{1.25}
\begin{array}{ll}
\chi_1^2=\eta_1^2=\zeta_1^2, ~&\text{if}~\lambda^2\neq 1\,,\\
\chi_1^2+\eta_1^2+\zeta_1^2=1, ~ &\text{if}~\lambda^2=1\,.
\end{array}
\right.
\ee
According to Fig.~\ref{fig:SmuSnuAndLambdavspsi}, $\lambda^2\!=\!1$ at two special points, the FM Heisenberg ($\psi\!=\!\pi$) and the dual FM Heisenberg ($\psi\!=\!1.6475\pi$) points. Therefore, at these points the coefficients $\chi_1$, $\eta_1$ and $\zeta_1$ span the two-dimensional $\mc{S}^2$ surface of the unit sphere, while elsewhere they can only take the values $\pm1$, i.e., they become Ising variables. So the LT method delivers the degeneracy expected for the FM and dual FM Heisenberg points. Away from these points the LT method gives eight GSs (except at the AF and dual AF Heisenberg points, see below). These eight states are related to each other by the symmetry operations of the dihedral group $\widetilde{\mathsf{D}}_2$ and time reversal.

\vspace*{-0.3cm}
\subsection{Classical ground states at special points ($\psi\!=\!\frac{3\pi}{2}$, $0$, $0.6475\pi$)}
\vspace*{-0.3cm}
\subsubsection{The FM Kitaev point ($\psi=3\pi/2$)}\label{sec:ClassGSsFMKitaev}
\vspace*{-0.3cm}
The FM Kitaev point is special in that the minimum of the eigenvalue spectrum of $\bs{\Lambda}_{\bf k}$ is achieved on the lines $l_x$, $l_y$ and $l_z$ of the BZ, and not just on the $\bs{\Gamma}$ point (see hexagon in Fig.~\ref{fig:PhaseDiagram}). 
This feature leads to a subextensive number of GSs, eight of which are the ones resulting from Eq.~(\ref{eq:ClassGSsIIAandIIB}) at $\lambda\!=\!0$. 

The whole set of GSs can arise from the subextensive GSs of the AF Kitaev point by simply applying the $K\mapsto-K$ duality transformations of Sec.~\ref{sec:KitaevDualities}. According to Fig.~\ref{fig:Dualities}\,(c), we must simply flip the sign of the $x$-, $y$- and $z$-components on every second site along each `xx', `yy' and `zz' line, respectively. The resulting $2^{3L}$ states are shown in Fig.~\ref{fig:IAIBandFMKitaev}\,(b). 
These states are again non-coplanar but the angle between any two NN spins is now $60^\circ$ and the spin components along a given Cartesian axis are correlated ferromagnetically along the corresponding `xx', `yy' or `zz' line.
As in the case of the AF Kitaev point, the $2^{3L}$ degeneracy is not accidental but stems from the Kitaev self-dualities of Sec.~\ref{sec:KitaevSelfDualities}, which amount to flipping the sign of any of the numbers $\chi_\ell$, $\eta_\ell$ or $\zeta_\ell$ in Fig.~\ref{fig:IAIBandFMKitaev}\,(b).

\vspace*{-0.3cm}
\subsubsection{The AF Heisenberg point ($\psi=0$) and its dual ($\psi=0.6475\pi$)}\label{sec:ClassGSsAFHeis}
\vspace*{-0.3cm}
The end points of the region IIA-IIB are also special in that the minimum of the eigenvalue spectrum of $\bs{\Lambda}_{\bf k}$ is achieved on the whole BZ and not just on the $\bs{\Gamma}$ point. For the KHAF this feature reflects the presence of an extensive manifold of GSs~\cite{HarrisKallinBerlinsky1992}. These states satisfy the condition that the total spin in each triangle vanishes, i.e.,
\be
{\bf S}_{{\bf R},\text{A}}+{\bf S}_{{\bf R},\text{B}}+{\bf S}_{{\bf R},\text{C}}=0\,,~~\forall {\bf R}\,.
\ee 
The resulting manifold includes an infinite subset of coplanar states as well as an infinite subset of non-coplanar states~\cite{HarrisKallinBerlinsky1992,Sachdev1992,Chubukov1993}. 
The eight uniform states resulting from Eq.~(\ref{eq:phaseI}) for $\chi_{\bf R}$, $\eta_{\bf R}$ and $\zeta_{\bf R}$ independent of ${\bf R}$, as well as the eight states of Eq.~(\ref{eq:ClassGSsIIAandIIB}) for $\lambda\!=\!2$  belong to this subset.
For $\psi\!=\!2\pi$, for example, the sublattices A, B and C approach the directions $[-2\chi_1, \eta_1, \zeta_1]$, $[\chi_1,-2\eta_1,\zeta_1]$ and $[\chi_1, \eta_1, -2\zeta_1]$, respectively, which corresponds to a uniform 120$^\circ$ coplanar configuration, see last inset graphic of Fig.~\ref{fig:SmuSnuAndLambdavspsi}. 
The situation for the dual AF Heisenberg point follows by the dualities $\mathsf{D}_A$, $\mathsf{D}_B$ and $\mathsf{D}_C$.

\begin{figure}[!b]
\includegraphics[width=0.45\linewidth]{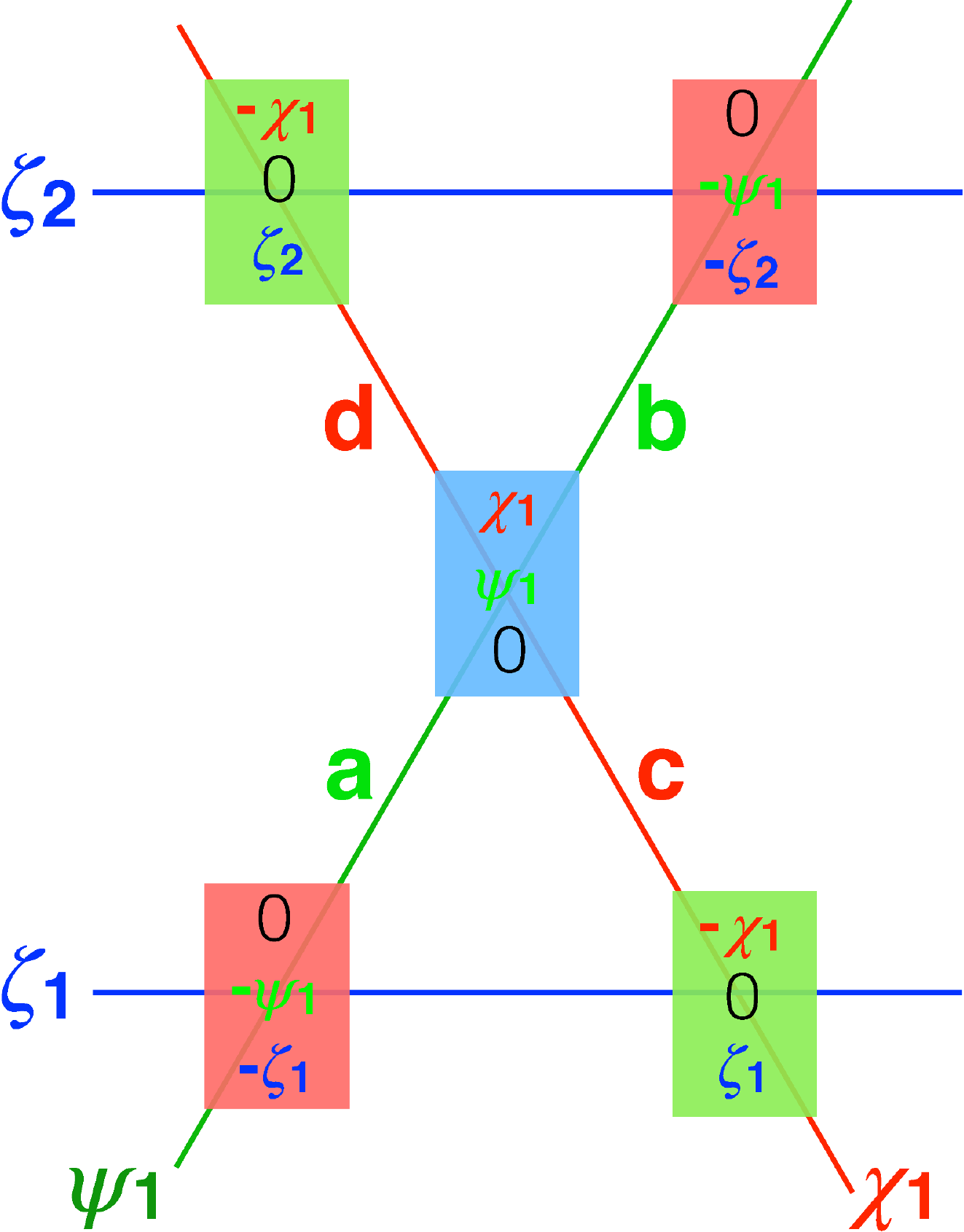}
\caption{\label{fig:RSPT} Minimal cluster mediating a coupling between NN Ising variables (here $\zeta_1$ and $\zeta_2$) in 4th-order real space perturbation theory.} 
\end{figure}

\vspace*{-0.3cm}
\section{Semiclassical analysis: Quantum order-by-disorder in the regions IA-IB}\label{sec:ObD}
\vspace*{-0.3cm}
We now move onto the quantum case, which we first try to approach by a semiclassical $1/S$ approach. In particular, let us first consider the question of the lifting of the accidental sub-extensive degeneracy inside the regions IA and IB (except the AF Kitaev point where the degeneracy is symmetry related as mentioned above). To address this question we follow the so-called real space perturbation theory (RSPT) approach~\cite{Lindgard1988,Long1989,Heinila1993,Mike2014,IoannisGammaModel,IoannisClassKitaev}.
In this approach one introduces a local axes frame along the classical spin directions of each spin site $i$, and then splits the Hamiltonian $\mc{H}$ into a diagonal part $\mc{H}_0$ that includes fluctuations in the local field, and a perturbation $\mc{V}\!=\!\mc{H}\!-\!\mc{H}_0$, which couples off-diagonal fluctuations on different sites (see App.~\ref{app:RSPT}).

In the present case, these fluctuations give rise to effective couplings between the Ising-like variables $\chi_\ell$, $\eta_\ell$ and $\zeta_\ell$ that parametrize the classical GSs of the regions IA and IB. As it turns out, the leading effective couplings appear in fourth-order perturbation theory, with lower order terms giving rise to global energy shifts. The fourth-order virtual processes involve five-site clusters, like the one shown in Fig.~\ref{fig:RSPT}. 
In this cluster, a coupling between the Ising-like variables $\zeta_1$ and $\zeta_2$ arises from virtual processes generated by the parts of $\mc{V}$ living on the bonds $a$, $b$, $c$ and $d$ of Fig.~\ref{fig:RSPT}. With the initial state being the same as the final state, the virtual process must involve either the bonds $\{a,  c\}$, or $\{a, b\}$, or $\{d, b\}$, or $\{d, c\}$. The individual contributions from these four types of processes (evaluated for $S\!=\!1/2$) are : 
\be
\renewcommand{\arraystretch}{1.45}
\begin{array}{ll}
\{a, b\} ~\text{or}~\{d, c\}: &
\delta E^{(4)}_{1} = -\frac{7}{288}\frac{J^2}{J+K}\zeta_1\zeta_2+\text{cst}\\ 
\{a, d\} ~\text{or}~\{b, c\}: & 
\delta E^{(4)}_{2} = +\frac{13}{576}\frac{J^2}{J+K}\zeta_1\zeta_2+\text{cst}
\end{array}
\ee
where `cst' denote global constants. Disregarding these constants (and the ones arising from second- and third-order processes) and adding all contributions gives the total effective coupling involving $\zeta_1$ and $\zeta_2$:
\be\label{eq:Jeff}
\delta E^{(4)}_{\zeta_1\zeta_2} = J_{\text{eff}}\zeta_1\zeta_2, 
~~J_{\text{eff}}=-L\times \frac{1}{288}\frac{J^2}{J+K}\,, 
\ee
where we have included a prefactor of $L$ to account for all five-site clusters connecting the lines $\zeta_1$ and $\zeta_2$. 
The fact that $J_{\text{eff}}\!\propto\!L$ guarantees a size extensive total energy in the effective model, and arise because the Ising variables account for the collective behaviour of whole lines of spins.

Similar effective couplings arise for all NN lines of any type, leading to the order-by-disorder effective Hamiltonian 
\be\label{eq:Heff}
\delta E^{(4)} = J_{\text{eff}}  \left(
\sum_{\langle \ell\ell'\rangle\in\text{`xx'}} \!\!\!\!\chi_\ell \chi_{\ell'}
+\!\!
\sum_{\langle \ell\ell'\rangle\in\text{`yy'}} \!\!\!\!
\eta_\ell \eta_{\ell'}
+\!\!
\sum_{\langle \ell\ell'\rangle\in\text{`zz'}} \!\!\!\!
\zeta_\ell \zeta_{\ell'} \right)\,.
\ee
In addition, the sign of $J_{\text{eff}}$ is negative everywhere inside the regions IA and IB of the classical phase diagram. Therefore, fourth-order virtual processes tend to align the Ising-like variables ferromagnetically. So the leading quantum fluctuations lift the $2^{3L}$ subextensive accidental degeneracy of the regions IA and IB (except for the AF Kitaev point) and select the eight states with $\chi_\ell\!=\!\chi_1$, $\eta_\ell\!=\!\eta_1$ and $\zeta_\ell\!=\!\zeta_1$ for all $\ell\!=\!1\cdots L$. These states are uniform (i.e., ${\bf S}_{{\bf R},\text{A}}\!=\!{\bf S}_{\text{A}}$, ${\bf S}_{{\bf R},\text{B}}\!=\!{\bf S}_{\text{B}}$ and ${\bf S}_{{\bf R},\text{C}}\!=\!{\bf S}_{\text{C}}$ for all ${\bf R}$) and globally coplanar.
The full quantum mechanical calculations presented below reveal that this leading order-by-disorder effect remains robust to all orders of perturbation theory, except near the AF Heisenberg point and its dual, where the physics is different. 

The tendency of quantum fluctuations to select the above eight uniform states has been noticed previously in the numerical results of Morita {\it et al}~\cite{Morita2018,Morita2019b}. The RSPT analysis presented here unveils the nature of the leading virtual processes responsible for this order by disorder effect, and the effective Hamiltonian of Eq.~(\ref{eq:Heff}) encapsulates this effect in a particularly simple form in terms of effective exchange couplings between emergent Ising-like variables.

An important comment for the AF Kitaev point is in order here. According to Eq.~(\ref{eq:Jeff}), the effective coupling $J_{\text{eff}}$ vanishes at $\psi\!=\!\pi/2$. This feature will in fact survive to all orders of perturbation theory, and is a direct consequence of the subextensive self-dualities of Sec.~\ref{sec:KitaevSelfDualities}, which prohibit any coupling between the Ising variables. 
These symmetries therefore explain naturally the reported~\cite{Morita2018,Morita2019b} absence of classical and quantum order by disorder at the Kitaev points. 
Still, this statement does not imply absence of long-range order in the thermodynamic limit, as a spontaneous breaking of the self-dualities is still possible at zero temperature~\cite{Batista2005}, see also discussion in Sec.~\ref{sec:Correlations}.

\begin{figure*}
\includegraphics[width=\textwidth]{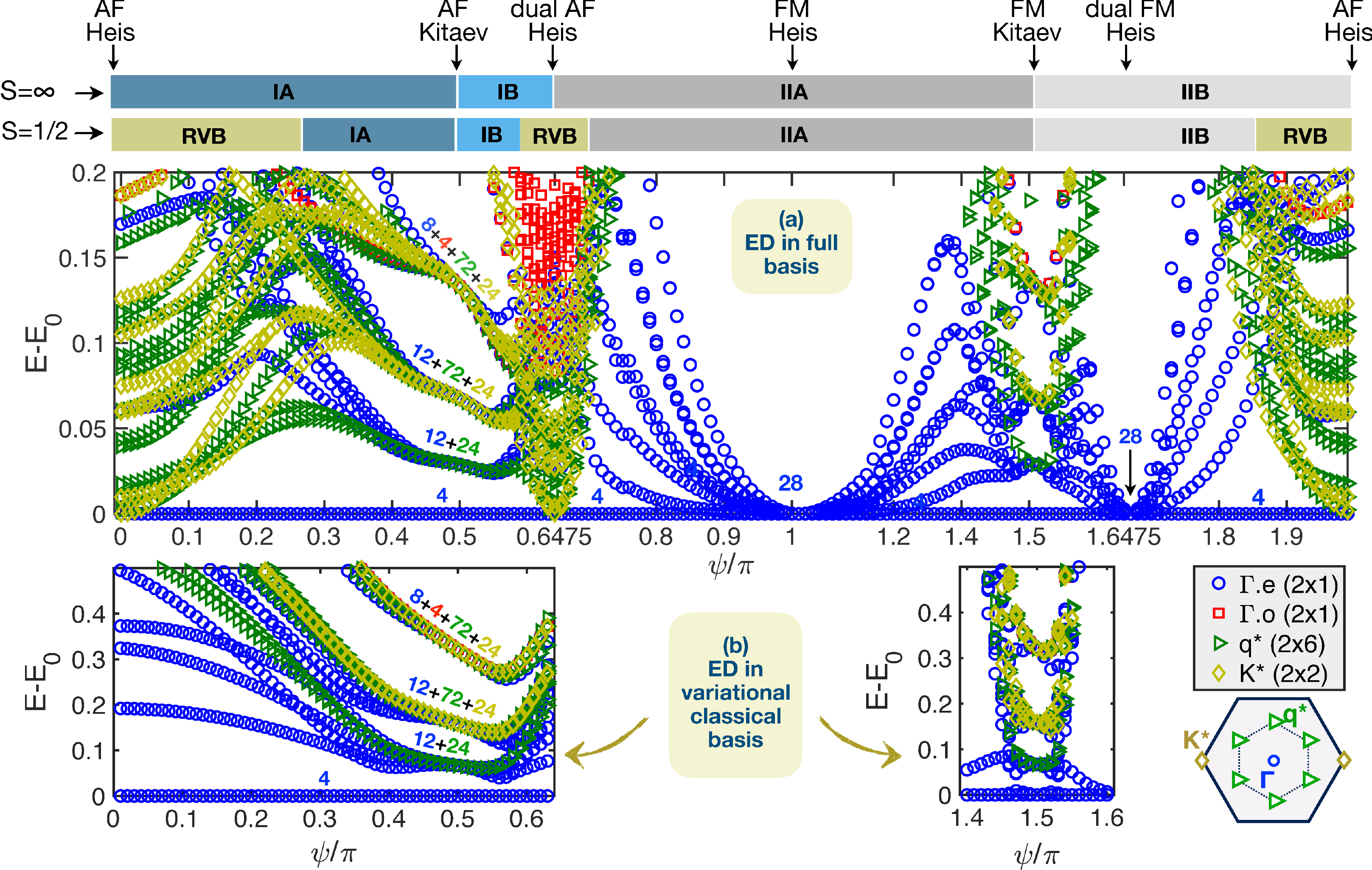}
\caption{The low-energy spectrum of the 27-site cluster measured from the GS energy $E_0(\psi)$. The symmetry sectors associated with the different symbols are shown in the right along with their dimensionalities (which here include an extra factor of 2 coming from time reversal).  Note that the horizontal axis is nonlinear in order to highlight the details of the spectrum near specific regions. 
(a) Upper panel: ED spectra in the full basis using the Lanczos algorithm. From the spectrum shown only the lowest 5 levels in each symmetry sector have converged to the requested accuracy of $10^{-12}$ in absolute energy.  
(b) Lower panels: 
ED spectra in the variational classical basis consisting of the 512 states of the region IA-IB (left panel), or the 512 states of the FM Kitaev point plus the 8 states of Eq.~(\ref{eq:ClassGSsIIAandIIB}) (right panel).}
\label{fig:spectrum27}
\end{figure*}

\vspace*{-0.3cm}
\section{Quantum spins $S\!=\!1/2$: Exact Diagonalization study}\label{sec:QuantumED}
\vspace*{-0.3cm}
We now turn to the study of the quantum $S=1/2$ version of the model (\ref{eq:Model}). To this end, we have performed exact diagonalizations (ED) on finite-size clusters with periodic boundary conditions, and have examined the symmetry structure of the low-energy spectrum and the GS spin-spin correlation patterns as we vary $\psi$. The interpretation of the resulting data is unveiled by contrasting the results from two independent types of ED calculations: i) ED in the full basis, and ii) ED in the restricted variational basis of the orthonormalized classical GSs of the regions IA-IB and IIA-IIB.  
Taken together, the results show that the qualitative semiclassical $1/S$ picture presented above remains robust down to $S\!=\!1/2$, except near the AF and dual AF Heisenberg points.

\vspace*{-0.3cm}
\subsection{Preliminaries}
\vspace*{-0.3cm}
The results presented here are obtained for a 24-site and a 27-site cluster, with spanning vectors $(\mathbf{T}_1,\mathbf{T}_2)=(\boldsymbol{a}_1+2\boldsymbol{a}_2,-3\boldsymbol{a}_1+2\boldsymbol{a}_2)$ and $(\mathbf{T}_1,\mathbf{T}_2)=(3\boldsymbol{a}_1,3\boldsymbol{a}_2)$, respectively.  
The 27-site cluster has the full point group $\mathsf{C}_{3v}$ symmetry of the infinite system, whereas the 24-site cluster has a lower ($\mathsf{C}_{2v}$) point group.
The nine allowed momenta of the 27-site cluster are shown in the bottom right corner of Fig.~\ref{fig:spectrum27}, and include the $\bs{\Gamma}$ point, the two corners $\pm {\bf K}$ of the Brillouin zone (labeled by ${\bf K}^\ast$), and six momenta inside the Brillouin zone (labeled by ${\bf q}^\ast$), which are related to each other by $\mathsf{C}_{3v}$. 
Similarly, the eight allowed momenta of the 24-site cluster are shown in the side inset of Fig.~\ref{fig:spectrum24full} and include the $\bs{\Gamma}$ point, a single ${\bf M}$ point, and three pairs $\pm {\bf q}_1$, $\pm {\bf q}_2$ and $\pm {\bf q}_3$ (not related to each other by any symmetry) inside the Brillouin zone.

In our exact diagonalizations we have implemented the symmetries under translations, real space inversion as well as spin inversion (i.e., global $\pi$-rotation around the ${\bf x}$-axis in spin space). 
The presented spectra therefore carry quantum numbers associated with the momentum ${\bf k}$, the parity (even or odd, denoted by `e' and `o', respectively) under real space inversion and the parity under spin inversion (even or odd, denoted by `Sze' and `Szo', respectively). For the 27-site cluster, the spin inversion leads to an extra twofold degeneracy due to Kramers theorem (i.e., the sectors `Sze' and `Szo' are degenerate).

The ED calculations in the restricted basis of classical GSs are performed as follows. We first generate the set of relevant classical states (depending on the region of $\psi$, see below) $|\alpha\rangle\!=\!\prod_{i=1}^{\mc{N}}|\bs{\Omega}_{i,\alpha}\rangle$, where
\be
|\bs{\Omega}_{i,\alpha}\rangle= \cos(\theta_{i,\alpha}/2) |\!\uparrow\rangle_i+e^{i \phi_{i,\alpha}} \sin(\theta_{i,\alpha}/2)|\!\downarrow\rangle_i\,.
\ee
Here $\alpha$ runs from one up to the number of classical GSs, $\mc{N}$ is the number of spins in the cluster, and $(\theta_{i,\alpha},\phi_{i,\alpha})$ are the spherical angles parametrizing the direction $\bs{\Omega}_{i,\alpha}$ of the $i$-th spin in the $\alpha$-th state. 
The restricted basis $\{|\alpha\rangle\}$ can then be orthonormalized using the overlap matrix $\mc{O}$, whose matrix elements are given by $\mc{O}_{\alpha\beta}\!=\!\langle\alpha|\beta\rangle$. We have checked numerically that the rank of this matrix equals its dimensionality (i.e., the states $\{|\alpha\rangle\}$ are linearly independent). 
The variational problem then reduces to diagonalizing the effective Hamiltonian
\be
\mc{H}_{\text{eff}}=\mc{O}^{-1/2}\mc{H}\mc{O}^{-1/2}\,,
\ee
where the matrix elements of $\mc{H}$ inside the basis $\{|\alpha\rangle\}$ can be found by splitting $\mc{H}$ into individual bond terms $\mc{H}_{ij}$ and
\be
\langle\alpha|\mc{H}_{ij}|\beta\rangle=\langle\bs{\Omega}_{i,\alpha},\bs{\Omega}_{j,\alpha}| \mc{H}_{ij}
|\bs{\Omega}_{i,\beta},\bs{\Omega}_{j,\beta}\rangle ~
\prod_{l \neq i,j} \langle\bs{\Omega}_{l,\alpha}|\bs{\Omega}_{l,\beta}\rangle\,.
\ee 

For the 27-site cluster ($L\!=\!3$) and for the region inside IA-IB, the restricted classical basis includes all $2^{3L}\!=\!512$ states of Fig.~\ref{fig:IAIBandFMKitaev}\,(a). 
For the region inside IIA-IIB, the basis includes the eight uniform states of Eq.~(\ref{eq:ClassGSsIIAandIIB}), as well as the 512 states of Fig.~\ref{fig:IAIBandFMKitaev}\,(b) which become relevant close to the FM Kitaev point.

\vspace*{-0.3cm}
\subsection{Low-energy spectra}
\vspace*{-0.3cm}
We are now ready to examine the ED spectra. The data from the two clusters are qualitatively consistent with each other, so here we focus on the  spectrum of the larger, 27-site cluster, shown in Fig.~\ref{fig:spectrum27} (the spectrum of the 24-site cluster is shown in Fig.~\ref{fig:spectrum24full} of App.~\ref{app:ED24site}).

First of all, due to the dualities $D_A$, $D_B$ or $D_C$, the spectrum in the regions IB and IIB can be mapped to the spectrum in the regions IA and IIA, respectively, by a rescaling [see Eq.~(\ref{eq:KtildeJtilde})]
\be\label{eq:IAtoIBmapping}
E(\widetilde{\psi})\!=\!\left[\frac{J^2+K^2}{\widetilde{J}^2+\widetilde{K}^2}\right]^{1/2}\!\!\!E(\psi)\!=\!\frac{E(\psi)}{\sqrt{3+2(\cos2\psi+\sin2\psi)}} \,.
\ee
Furthermore, since the duality transformations are uniform and do not break real space inversion, the  momentum and parity quantum numbers of the spectra are retained in the mapping. 

Likewise, the spectra at the isolated points $\psi\!=\!\pm\pi/2$ are identical to each other, as can be clearly seen in Fig.~\ref{fig:spectrum27}. This is due to the duality that maps $K\!\to\!-K$, mentioned in Sec.~\ref{sec:KitaevDualities}. 

\begin{figure*}[!t]
\includegraphics[width=\textwidth]{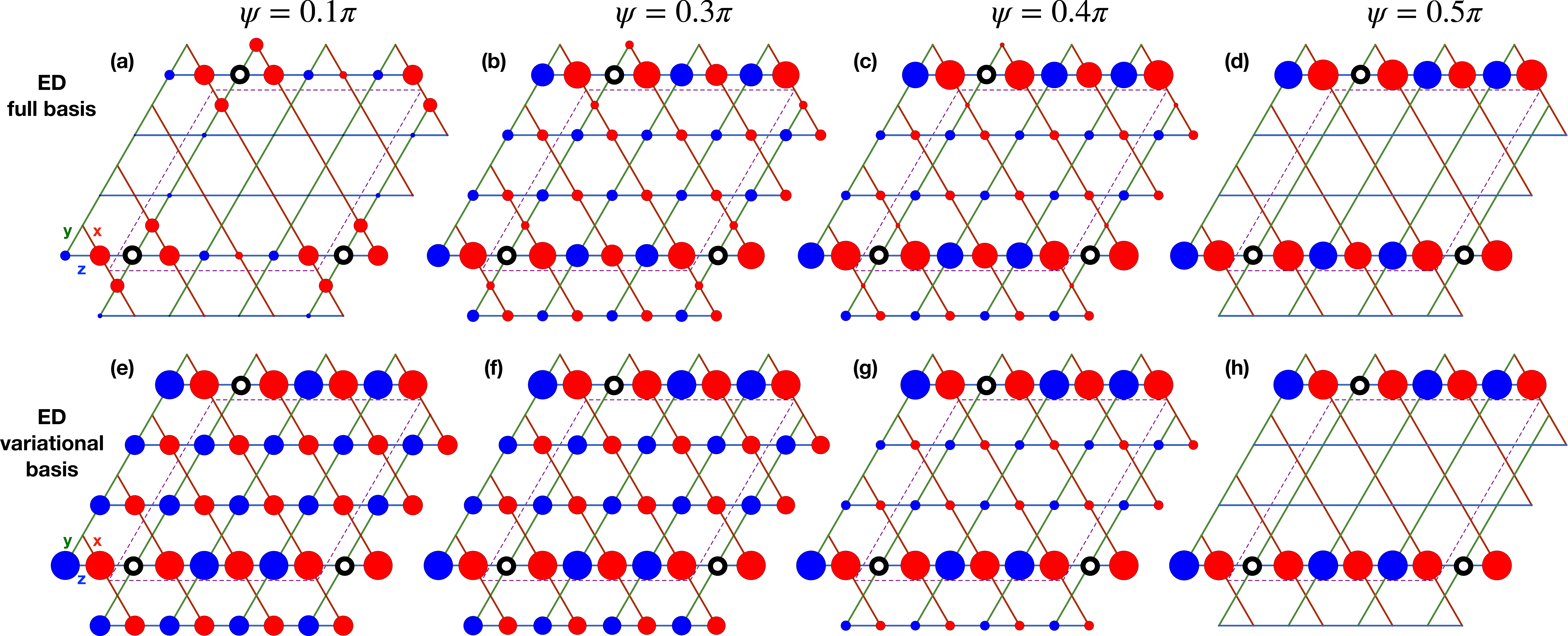}
\caption{Real-space pattern of GS spin-spin correlations $\langle S_i^zS_j^z\rangle$ in the 27-site cluster (enclosed by dashed lines) for four representative points of the region IA (the corresponding data for $\langle S_i^xS_j^x\rangle$ and $\langle S_i^yS_j^y\rangle$ can be obtained from $\langle S_i^zS_j^z\rangle$ by three-fold rotations). 
The two rows of panels show data obtained using ED in the full basis (a-d) and ED in the restricted classical basis (e-h) of the 512 states of Fig.~\ref{fig:IAIBandFMKitaev}\,(a). The black open circle at each panel denotes the reference site $i$. Positive (negative) correlations are shown by filled blue (filled red) circles, whose radius scales with the magnitude of the correlation. The strongest correlation, corresponding to NN sites in (d), is about -0.137; the classical value is $-S^2/2\!=\!-0.125$, see Eq.~(\ref{eq:s1zsjzzeta1zeta2}). Incidentally, this number is very close to the exact result -0.13 for the NN correlations in the Kitaev Honeycomb model in the thermodynamic limit~\cite{Baskaran2007}.}\label{fig:correlation-1} 
\end{figure*} 

Second, apart from the vicinities of the AF and dual AF Heisenberg points (see below), the spectra in the full and the restricted basis are qualitatively the same, both in the multiplicities of the levels and the symmetry quantum numbers, up to a relatively high excitation energy. This is the first strong spectral evidence that the semiclassical picture survives down to $S\!=\!1/2$, except near the AF and dual AF Heisenberg points. 

Let us examine some individual subregions, starting with the vicinity of the AF Kitaev point. Here we find a characteristic sequence of highly degenerate levels consisting of both zero- and finite-momenta states. 
Exactly at $\psi\!=\!\pi/2$, we find, above the fourfold degenerate GS (decomposing into $2\bs{\Gamma}.\text{e}$), a 36-fold first-excited level (decomposing into $6\bs{\Gamma}.\text{e}\oplus2{\bf q}^\ast$), followed by a 108-fold level (decomposing into $6\bs{\Gamma}.\text{e}\oplus6{\bf q}^\ast\oplus6{\bf K}^\ast$), followed by another 108-fold level (decomposing into $4\bs{\Gamma}.\text{e}\oplus2\bs{\Gamma}.\text{o}\oplus6{\bf q}^\ast\oplus6{\bf K}^\ast$), and so on. 
The qualitative agreement~\footnote{Note that the excitation energies in Fig.~\ref{fig:spectrum24full}\,(a) are generally pushed down compared to the ones in Fig.~\ref{fig:spectrum24full}\,(b), due to the repulsion from the states not belonging to the classical manifold.} with the variational ED data shown in the left panel of Fig.~\ref{fig:spectrum27}\,(b) demonstrates that this characteristic pattern is a manifestation of expected tunnelling among the 512 classical GSs due to the finite value of $S\!=\!1/2$ (the tunnelling goes to zero and the levels collapse to each other in the limit $S\!\to\!\infty$). Moreover, the observed high multiplicities of the levels arise from the fact that the Kitaev point features a nontrivial symmetry group, consisting of a subextensive number of operations (512 for the 27-site cluster), which in turn leads to irreducible representations of very high dimensionality. 

For completeness, we have carried out an independent symmetry analysis of the 512 classical GSs of the 27-site cluster and have found the following symmetry decomposition 
\be
\text{512 states of IA-IB} \to \text 36\bs{\Gamma}.\text{e}\oplus4\bs{\Gamma}.\text{o}\oplus28{\bf q}^\ast\oplus24{\bf K}^\ast\,.
\ee
From these 512 states, only 256 are visible in the energy range shown in the full basis ED results of Fig.~\ref{fig:spectrum27}\,(a), exactly half from each sector. The remaining half reside at higher energies, and some of them mix with states outside the classical manifold.

The FM Kitaev point shows an identical pattern of highly degenerate levels, but these levels go up in energy very fast as we depart from $\psi\!=\!3\pi/2$, in contrast to what happens around $\psi\!=\!\pi/2$. This qualitative difference originates from the fact that the respective 512 states are classical GSs only at $\psi\!=\!3\pi/2$, and away from this point these get replaced by the 8 states of Eq.~(\ref{eq:ClassGSsIIAandIIB}). Moreover, this difference is captured by the variational energy spectra shown in right panel of Fig.~\ref{fig:spectrum27}\,(b).

Next, in the regions extending roughly between $\psi\!\sim\!0.7\pi$ and $1.45\pi$ and between $\psi\!\sim\!1.55\pi$ and $1.85\pi$, the low-energy spectrum consists entirely of zero-momentum states which are well separated in energy from finite-momentum states. Furthermore, all low-lying states have even parity with respect to real space inversion. These features suggest that the system does not break the translation and inversion symmetry in these regions, in agreement with the predictions of the semiclassical analysis (given that the states of Eq.~(\ref{eq:ClassGSsIIAandIIB}) as well as the 28 states of the $S\!=\!27/2$ GS manifold of the FM Heisenberg point are all uniform and even). 

Finally, let us discuss what happens in the vicinities of the AF and dual AF Heisenberg points. Figure~\ref{fig:spectrum27}\,(a) shows a rapid rearrangement of the low-energy spectrum, with a very large number of states coming down in energy near these points, leading to a very dense spectrum in all possible momentum sectors. 
These states are clearly not related to the 512 states around the AF Kitaev point, as can be seen by a direct comparison to the left panel of Fig.~\ref{fig:spectrum27}\,(b). 
The dense excitations are in fact not unexpected, as it is well known that the low-lying spectrum of the KHAF features an extensive number of low-lying excitations, in both the singlet and the higher spin sectors~\cite{Lecheminant1997,Waldtmann1998,Mila1998,Sindzingre2009,Rousochatzakis2008,Laeuchli2019}. 
It is further known that the origin of these low-lying states is actually {\it not} classical~\cite{Rousochatzakis2008,Gotze2011} but strongly quantum~\cite{Mila1998,ZengElser95,MambriniMila2000,Misguich03,Poilblanc2010,Ralko2018,Laeuchli2019}.

The competition between the quantum low-lying states of the KHAF and the semiclassical states favoured by the Kitaev Hamiltonian (in particular, the eight states stabilized via the order-by-disorder mechanism in IA and IB, see Sec.~\ref{sec:ObD}) appears to give rise to a phase transition between the two, at some critical point $K/J$ (and a similar transition around the dual point $\psi\!=\!0.6475\pi$). 
A rough estimate for this point can be deduced from the observation that the rapid rearrangement of the spectrum happens around $\psi\!\sim\!0.25\pi$ in both 27- and 24-site clusters. The spin-spin correlation data from these two clusters alone give a similar rough estimate, as we show in the next section. 
This issue requires further discussion, however, and we will return to it in Sec.~\ref{sec:Discussion}.

\vspace*{-0.3cm}
\subsection{Spin-spin correlations}\label{sec:Correlations}
\vspace*{-0.3cm}
Further insights arise by examining the GS spin-spin correlations. We will focus on results obtained for the 27-site cluster, and we will present correlations of the type $\langle S_i^zS_j^z\rangle$, since the corresponding patterns for $\langle S_i^xS_j^x\rangle$ and $\langle S_i^yS_j^y\rangle$ arise by threefold rotations.

\begin{figure*}[!t]
\includegraphics[width=\textwidth]{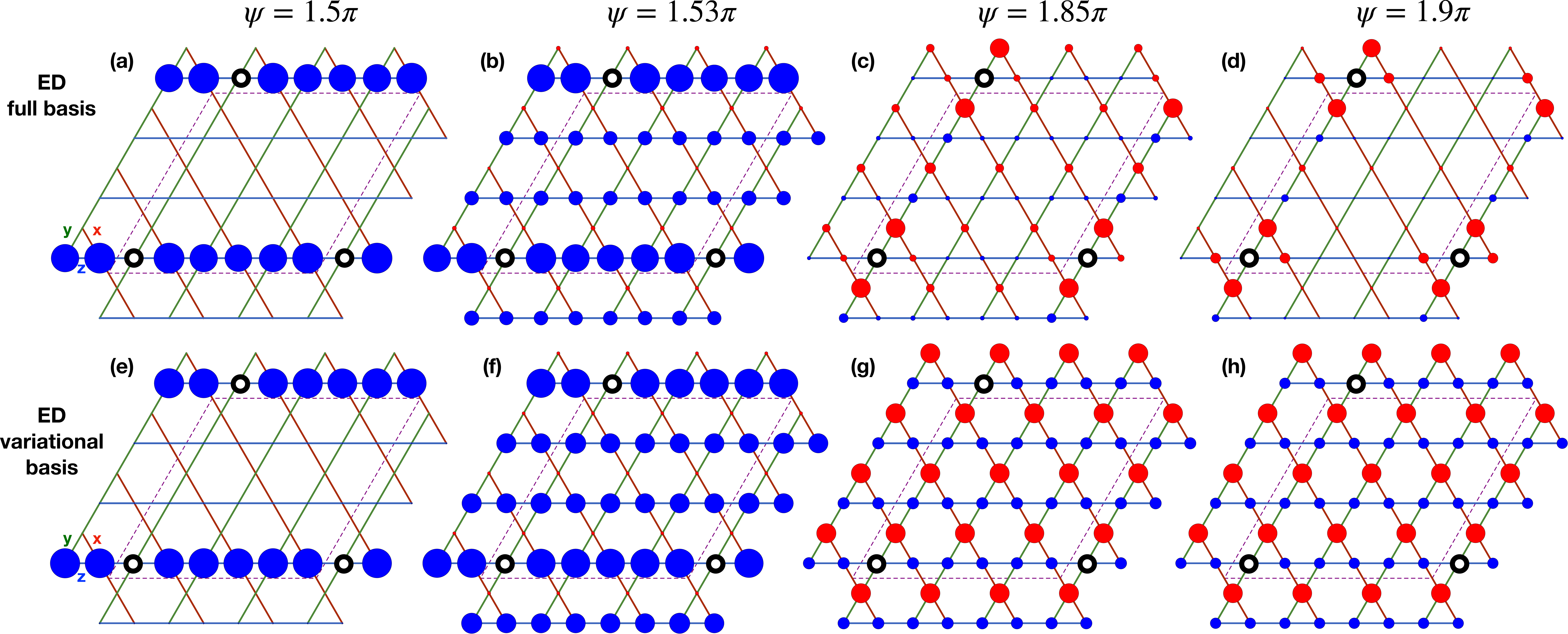}
\caption{Real-space pattern of spin-spin correlations $\langle S_i^z S_j^z\rangle$ for the 27-site cluster (enclosed by dashed lines) for four representative points of the region IIB. The two rows of panels show data obtained from ED in the full basis (a-d) and ED in the restricted classical basis of the 512 states of the FM Kitaev point plus the 8 states of Eq.~(\ref{eq:ClassGSsIIAandIIB}) (e-h). The black open circle at each panel denotes the reference site $i$. Positive (negative) correlations are shown by filled blue (filled red) circles, whose radius scales with the magnitude of the correlation. The strongest correlation, corresponding to NN sites in (a), is about 0.137; the classical value is $S^2/2\!=\!0.125$, see Fig.~\ref{fig:IAIBandFMKitaev}\,(b).}\label{fig:correlation-2} 
\end{figure*}

\vspace*{-0.3cm}
\subsubsection{Regions IA-IB}
\vspace*{-0.3cm}
We first examine the correlation patterns at four representative points inside the region IA, $\psi\!=\!0.1\pi$, $0.3\pi$, $0.4\pi$ and $0.5\pi$, see Fig.~\ref{fig:correlation-1}\,(a-d). The corresponding correlations inside IB can be deduced via the dualities $\mathsf{D}_A$, $\mathsf{D}_B$, and $\mathsf{D}_C$. 

We begin by analyzing the results for the Kitaev point [Fig.~\ref{fig:correlation-1}\,(d)]. Here we observe a strongly anisotropic profile with correlations being nonzero only along the single $zz$-line containing the reference site $i$ (black open circle) and alternating in sign from one site to the next.
This characteristic pattern is similar to what happens in the compass model in the square lattice~\cite{Dorier2005,Nussinov2015}, and its origin is the following. 
According to Fig.~\ref{fig:IAIBandFMKitaev}\,(a), for any classical GS, the correlations $\langle S_i^zS_j^z\rangle$ are nonzero along all $zz$ lines and not just along a single $zz$ line. 
Take, for example, two spin sites in Fig.~\ref{fig:IAIBandFMKitaev}\,(a) one (${\bf S}_i$) sitting on the horizontal line labeled by $\zeta_1$ and the other (${\bf S}_j$) on the line labeled by $\zeta_2$. From the general form of the GSs, we find that the classical $S_i^zS_j^z$ correlation is given by:
\be
i~\text{on}~\zeta_1\text{-line},~ j~\text{on}~\zeta_2\text{-line}\!: \left(S_i^zS_j^z\right)_{\text{cl}} \!\!\!=\! (-1)^{d_{ij}} (S^2/2)\zeta_1 \zeta_2\,,
\ee
where $d_{ij}$ is the distance between $i$ and $j$. 
So, for fixed $i$ and $j$, the overall sign of the correlation depends on the relative signs of $\zeta_1$ and $\zeta_2$. As such, it is positive for half the states in the classical GS manifold, and negative for the other half.  
Now, the quantum GS of the 27-site cluster at $\psi\!=\!\pi/2$ is equal, to a very good approximation, to a symmetric superposition of all 512 classical states of Fig.~\ref{fig:IAIBandFMKitaev}\,(a). This is corroborated by the agreement between Fig.~\ref{fig:spectrum27}\,(a) and (b, left panel), as well as the agreement between Fig.~\ref{fig:correlation-1}\,(d) and (h) for the correlations (Note, in particular, the striking agreement in the NN correlations $\langle S_i^zS_j^z\rangle\!=\!-0.136999$ from ED in the full basis and $-0.133779$ from ED in the variational basis).   
As a result, the above type of correlations (i.e., between spins belonging to different $zz$ lines) average out in this superposition. 
This cancellation does not occur when the two sites sit on the same $zz$ line, since in that case
\be\label{eq:s1zsjzzeta1zeta2}
i \,\& j~\text{on}~\zeta_1\text{-line}: ~ \left(S_i^zS_j^z\right)_{\text{cl}} \!=\! -(S^2/2)\zeta_1^2=-S^2/2\,,
\ee
the same for all members of the classical GS manifold.

It is important to emphasize that the absence of correlations between different $zz$ lines is only a finite-size (or finite-$S$) effect, as the system can still develop long-range order in the limit $\mc{N}\!\to\!\infty$ (or $S\!\to\!\infty$, respectively) at zero temperature~\cite{Batista2005}. The actual correlation pattern will then depend on the particular member of the manifold that is selected spontaneously. This is similar to what happens in the square lattice compass model at zero temperature~\cite{Dorier2005}.

\begin{figure*}[!t]
\centering
\includegraphics[width=\linewidth]{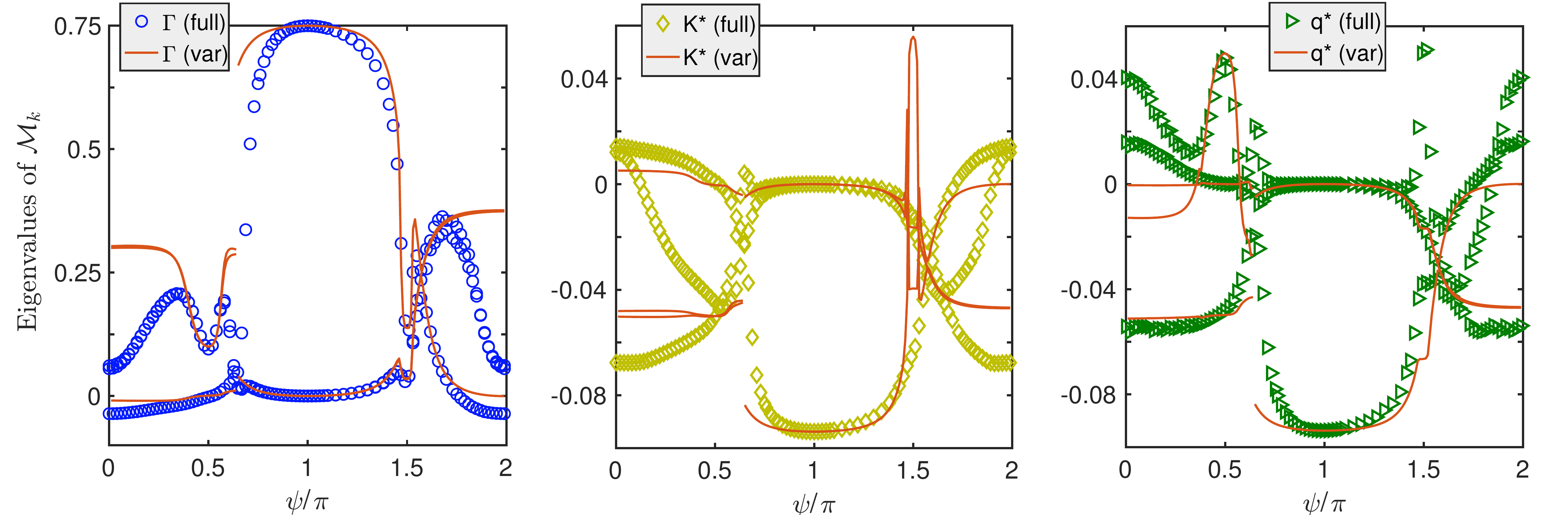}
\caption{Evolution of the eigenvalues of the correlation matrix $\mc{M}_{\bf k}$ of Eq.~(\ref{eq:CorMatrix}), evaluated in the GS of the 27-site cluster and for all allowed momenta of the cluster. There are two sets of data, one from ED in the full basis (symbols) and the other from ED in the orthonormalized classical basis of each region. The maximum eigenvalue in (a) at $\psi\!=\!\pi$ is 0.75 which is consistent with the value $3S^2$ expected for the FM GS [see definition of $\mc{M}_{\bf k}$ in Eq.~(\ref{eq:CorMatrix})].}
\label{fig:slength27}
\end{figure*}

Moving away from the AF Kitaev point, the correlations on $zz$ lines not including the reference site become nonzero even for finite-sizes, as can be seen in Figs.~\ref{fig:correlation-1}\,(c, g), (b, f) and (e). This happens because, away from the Kitaev point, the 512 states are classically degenerate by accident and not by symmetry. As such, the degeneracy is lifted by quantum fluctuations, leading to the eight uniform states discussed in Sec.~\ref{sec:ObD}. The fact that the pattern seen in Figs.~\ref{fig:correlation-1}\,(c, g), (b, f) and (e) is qualitatively consistent with that of the eight uniform states is therefore a numerical confirmation of the order by disorder effect of Sec.~\ref{sec:ObD}. Similar numerical confirmation has been reported by Morita {\it et al}~\cite{Morita2018,Morita2019b}.

We should further note that the spatially anisotropic profiles shown in Figs.~\ref{fig:correlation-1}\,(c,g) and (b,f), i.e., the fact that the correlations along the $zz$ line of the reference site are much stronger than those on the remaining $zz$ lines, reflects the presence of the remaining members of the classical manifold at low energies, at an energy scale $\propto\!J_{\text{eff}}$ given by Eq.~(\ref{eq:Jeff}). Given that the latter grows with $L$, one expects that the influence of these remaining members will diminish with $L$, and the correlations to become eventually uniform in strength throughout the bulk of the system for $L\!\to\!\infty$. In other words, the spatial anisotropy in the profiles of Figs.~\ref{fig:correlation-1}\,(c,g) and (b,f) is a finite-size effect.

The pattern seen in Fig.~\ref{fig:correlation-1}\,(a) for $\psi\!=\!0.1\pi$ is different from this picture [as also seen by the contrast with Fig.~\ref{fig:correlation-1}\,(e)], and is more consistent with the AF short-range correlations expected for the KHAF. The crossover to the long-range pattern begins around $\psi\!\sim\!0.25\pi$ as seen from the spectra reorganization discussed above, as well as in the evolution of the eigenvalues of the correlation matrix discussed in Sec.~\ref{sec:SF}.

\vspace*{-0.3cm}
\subsubsection{Regions IIA-IIB}
\vspace*{-0.3cm}
Next, we examine the correlation patterns of $\langle S_i^z S_j^z\rangle$ at four representative points inside the region IIB, $\psi\!=\!1.5\pi$, $1.53\pi$, $1.85\pi$ and $1.9\pi$, see Fig.~\ref{fig:correlation-2}\,(a-d). The corresponding correlations inside IIA can be deduced via the dualities $\mathsf{D}_A$, $\mathsf{D}_B$, and $\mathsf{D}_C$. 
The situation at $\psi\!=\!1.5\pi$ [Fig.~\ref{fig:correlation-2}\,(a)] is analogous to that at $\psi\!=\!\pi/2$, the only difference being that the amplitudes are FM along the $zz$ line of the reference site.  
The correlation profiles away from the FM Kitaev point [Figs.~\ref{fig:correlation-2}\,(b,f), (c,g) and (h)] are all consistent with the eight uniform states of Eq.~(\ref{eq:ClassGSsIIAandIIB}). In particular, the sign of the NN correlations switches from positive (FM) to negative from $1.53\pi$ to $1.85\pi$, in  full agreement with the classical picture, which predicts a sign change at $\psi\!=\!1.5695\pi$. 
Finally, the contrast between the patterns of Figs.~\ref{fig:correlation-2}\,(d) and (h) highlights once again the  RVB physics in the vicinity of the AF Heisenberg point (note, in particular, the similar profiles in Figs.~\ref{fig:correlation-1}\,(a) and  \ref{fig:correlation-2}\,(d)).

\vspace*{-0.3cm}
\subsubsection{Spin-spin correlation matrix}\label{sec:SF}
\vspace*{-0.3cm}
To shed further light into the crossover between the short-range physics in the vicinity of the AF Heisenberg point and the long-range ordering favoured by the Kitaev interactions we examine the evolution of the eigenvalues of the GS spin-spin correlation matrix $\mc{M}_{\bf k}$, defined as 
\be\label{eq:CorMatrix}
\mc{M}^{\mu\nu}_{\bf k}=\frac{1}{\mc{N}/3-1}\sum_{\bf R \neq 0} e^{-i{\bf k}\cdot{\bf R}} \langle {\bf S}_{0,\mu}\cdot{\bf S}_{{\bf R},\nu}\rangle\,,
\ee
where $\mu$, $\nu$ run over the three sublattices A, B and C of the kagome lattice, and ${\bf k}$ belongs to the set of the allowed momenta of the given cluster. 

 Figure~\ref{fig:slength27} shows the evolution of the eigenvalues of $\mc{M}_{\bf k}$ for the 27-site cluster, for all allowed ${\bf k}$ points and for two sets of GSs, one obtained in the full basis (symbols) and one in the restricted classical basis (lines). Quite generally, the maximum eigenvalue of $\mc{M}_{\bs{\Gamma}}$ is much larger than that of $\mc{M}_{{\bf q}^\ast}$ and $\mc{M}_{{\bf K}^\ast}$.
But most importantly, the comparison between the ED data in the full basis and the ones in the variational classical basis show almost perfect agreement everywhere, except in the vicinities of the AF and dual AF Heisenberg points. This is consistent with the general picture obtained from the low-energy spectra that the semiclassical physics remains robust down to $S\!=\!1/2$, as long as we are sufficiently away from the AF and dual AF Heisenberg points. Furthermore, the crossover between the RVB physics of the KHAF and the long-range ordering favoured by the Kitaev couplings seems to occur again around $\psi\!\sim\!0.25\pi$, mirroring the onset of the low-lying spectral rearrangement discussed above.

\vspace*{-0.3cm}
\section{Discussion}\label{sec:Discussion}
\begin{figure}[!t]
\centering
\includegraphics[width=\linewidth]{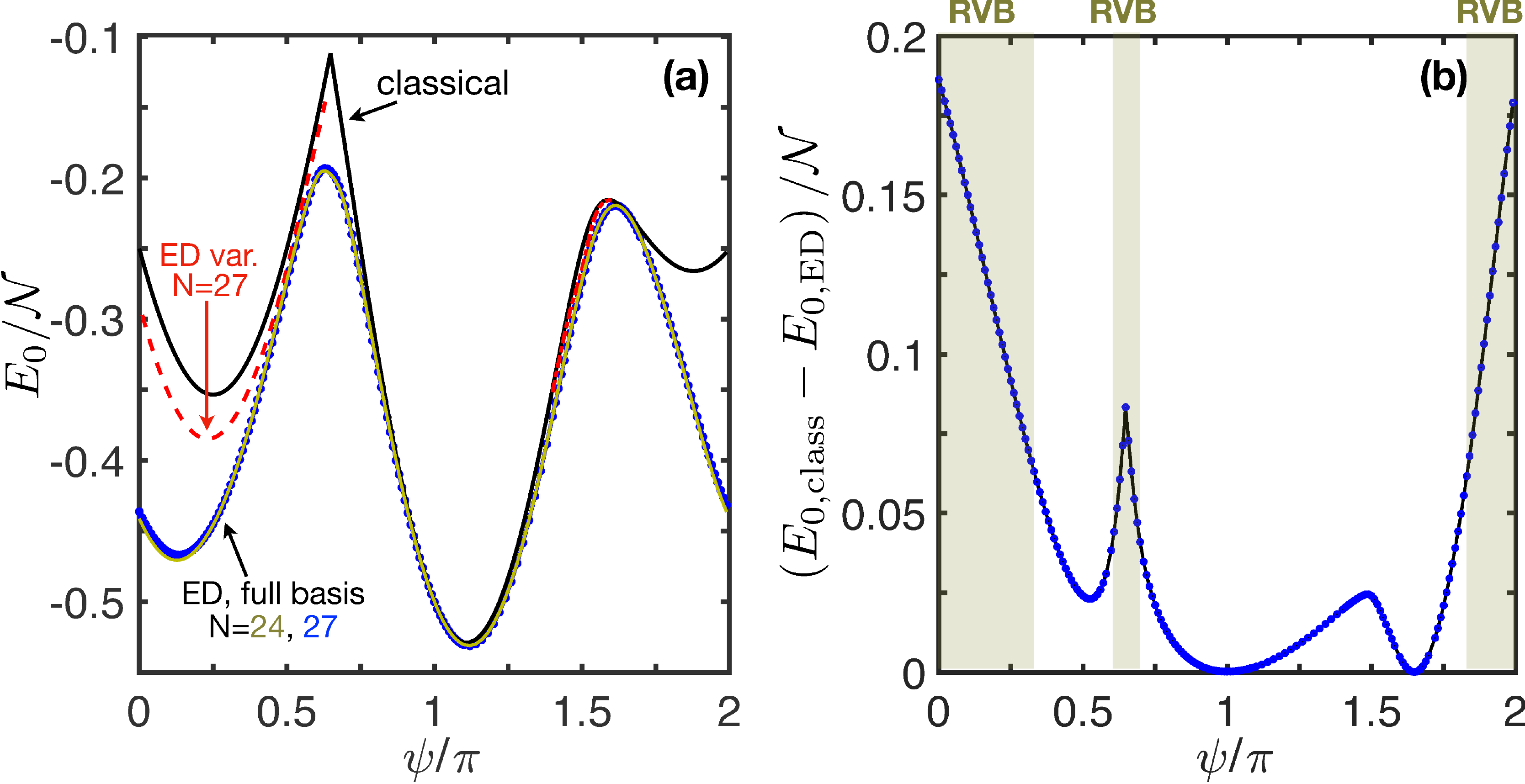}
\caption{(a) Evolution of ground state energy per site, $E_0/\mc{N}$, with $\psi$. The different curves correspond to: the classical result (black line), results from ED in the full basis of the 24- and the 27-site clusters (blue and dark yellow), and results from ED in the variational basis of the 27-site cluster (dashed red). 
(b) The difference between the classical energy per site and the energy obtained from ED in the full basis of the 27-site cluster.}
\label{fig:GSEnergy}
\end{figure}

\vspace*{-0.3cm}
In conjunction with previously known results~\cite{Kimchi2014,Morita2018,Morita2019b}, the present study concludes a rather consistent picture for the zero-temperature physics of the spin-1/2 Heisenberg-Kitaev model on the kagome lattice. 
Our ED results show a clear quantum-classical crossover from the RVB physics of the Heisenberg antiferromagnet to the regime stabilized by Kitaev anisotropy. The striking agreement between the ED results in the full basis and in the restricted classical GS basis in the Kitaev regime shows that this regime has a strong semiclassical character. 
By this we do not only mean that the GSs in this regime are qualitatively captured by the semiclassical limit, but also that the quantum (e.g., spin-wave) corrections are largely quenched. This is based on the fact that the ED in the restricted basis captures only the quantum tunneling between the different members of the basis, and not the fluctuation corrections from states outside the basis. 
The strong quenching of these corrections is essentially a manifestation of the large spin gaps generated by the Kitaev anisotropy. 
This aspect is also demonstrated in Fig.~\ref{fig:GSEnergy}, which shows that, inside the Kitaev regime, the classical ground state energies are very close to the corresponding quantum energies found from ED on finite sizes.

On a broader perspective, this crossover originates in the qualitatively different degree of frustration between the AF Heisenberg model on one hand and the Kitaev model on the other. Indeed, unlike what happens in the honeycomb lattice~\cite{Kitaev2006}, the Kitaev anisotropy in the kagome lattice does not come with any {\it local} conservation laws (explicit or emergent), but only with one-dimensional symmetries, much like what happens in the compass model on the square lattice~\cite{Doucot2005,Dorier2005}.

The presence of the one-dimensional symmetries has not been recognized before and explains naturally a number of features, including the reported absence of classical and quantum order by disorder at the AF Kitaev point in finite-size calculations~\cite{Morita2018,Morita2019b}. 
We re-emphasize here that this absence of order by disorder does not imply absence of long-range order. Such an order is allowed by the generalized Elitzur's theorem by Batista and Nussinov~\cite{Batista2005} at zero temperature, but its diagnosis in finite-size calculations is not straightforward as discussed in Sec.~\ref{sec:Correlations}.

Importantly, the above one-dimensional operations cease to be symmetries as soon as we depart from the Kitaev points. As a result, the subextensive classical GS degeneracy inside the region IA-IB of the parameter space (away from $\psi\!=\!0$ and $\pi/2$) is accidental, and is therefore lifted by fluctuations, a result that has been highlighted by Morita {\it et al}~\cite{Morita2018}. The perturbative analysis of Sec.~\ref{sec:ObD} provides a rather intuitive picture for this order by disorder effect, in terms of emergent interactions between collective, Ising-like variables describing whole lines of spins. This effective description leads naturally to the uniform coplanar states found in \cite{Morita2018} and also revealed in our spin-spin correlation results.

Finally, we return to the important question of whether the above quantum-classical crossover will show up in the thermodynamic limit with a phase transition at a nonzero value of $K$. Namely, whether the RVB phase of the KHAF will survive in an extended region around $K\!=\!0$. Our ED results for the low-energy spectra and spin-spin correlations show a qualitative change around $|K|\!\sim\!J$ for both 24- and 27-size clusters. By itself, this observation is not conclusive and a more systematic system-size dependence is needed to address this question.
We believe however that energetic considerations alone provide a clear insight that there will be a transition at a nonzero value of $K$, irrespective of the nature of the GS at $K\!=\!0$. On one hand, the strong quenching of fluctuation corrections mentioned above implies a weak renormalization of the GS energy in the semiclassical regime. Moreover, some of the classical GSs of the region IA-IB remain members of the classical GS manifold of the KHAF. On the other hand, we know that the low-energy states in the vicinity of $K\!=\!0$ are not related to the semiclassical limit~\cite{Rousochatzakis2008,Gotze2011} and have a strong quantum character~\cite{Mila1998,ZengElser95,MambriniMila2000,Misguich03,Poilblanc2010,Ralko2018,Laeuchli2019}. 
This is explicitly demonstrated in Fig.~\ref{fig:GSEnergy}\,(b) which shows that the classical states become energetically very costly around the KHAF and its dual. The classical orders of the regime IA or IIB must therefore overcome this large quantum energy cost before they become ground states deep inside the Kitaev regime. 
Importantly, this large energy cost is not related to the energy gap (if any) above the quantum ground state of the KHAF.  
Hence, the RVB phase of the KHAF will survive in a finite parameter range irrespective of whether this phase has a gap or not, and irrespective of the actual nature of the ground state. 

\vspace*{0.3cm} 
\noindent{\it Acknowledgments:} 
This work was supported by the U.S. Department of Energy, Office of Science, Basic Energy Sciences under Award No. DE-SC0018056. We also acknowledge the support of the Minnesota Supercomputing Institute (MSI) at the University of Minnesota.

\appendix
\titleformat{\section}{\normalfont\bfseries\filcenter}{Appendix~\thesection:}{0.25em}{}

\begin{figure*}[!t]
\includegraphics[width=\textwidth]{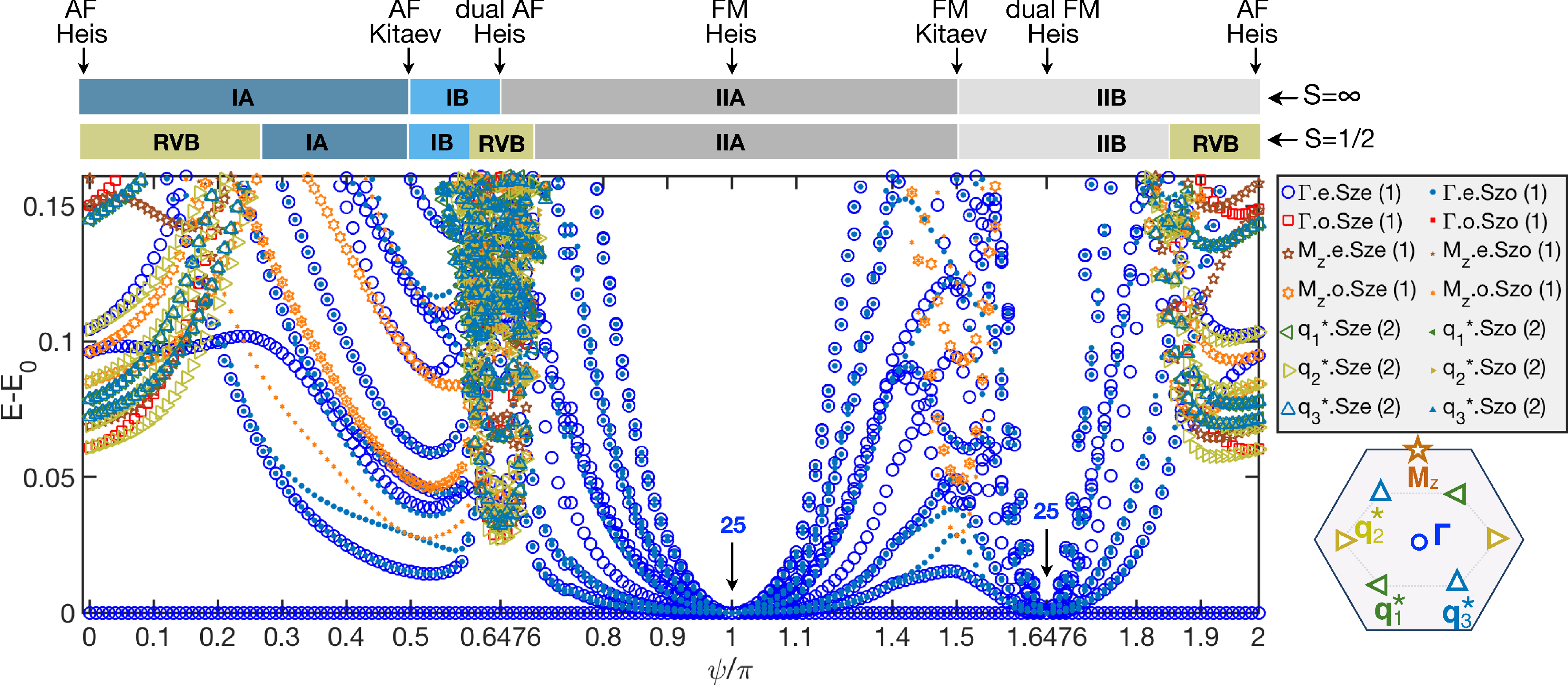}
\caption{The low-energy spectrum of the 24-site cluster measured from the GS energy $E_0(\psi)$. The symmetry sectors associated with the different symbols are shown on the right along with their dimensionalities. Note that the horizontal axis is nonlinear in order to highlight the details of the spectrum near specific regions. 
The data shown are obtained from ED in the full basis using the Lanczos algorithm. From the spectrum shown only the lowest 5 levels in each symmetry sector have converged to the requested accuracy of $10^{-12}$ in absolute energy.}\label{fig:spectrum24full}
\end{figure*}

\begin{figure}[!h]
\centering
\includegraphics[width=\linewidth]{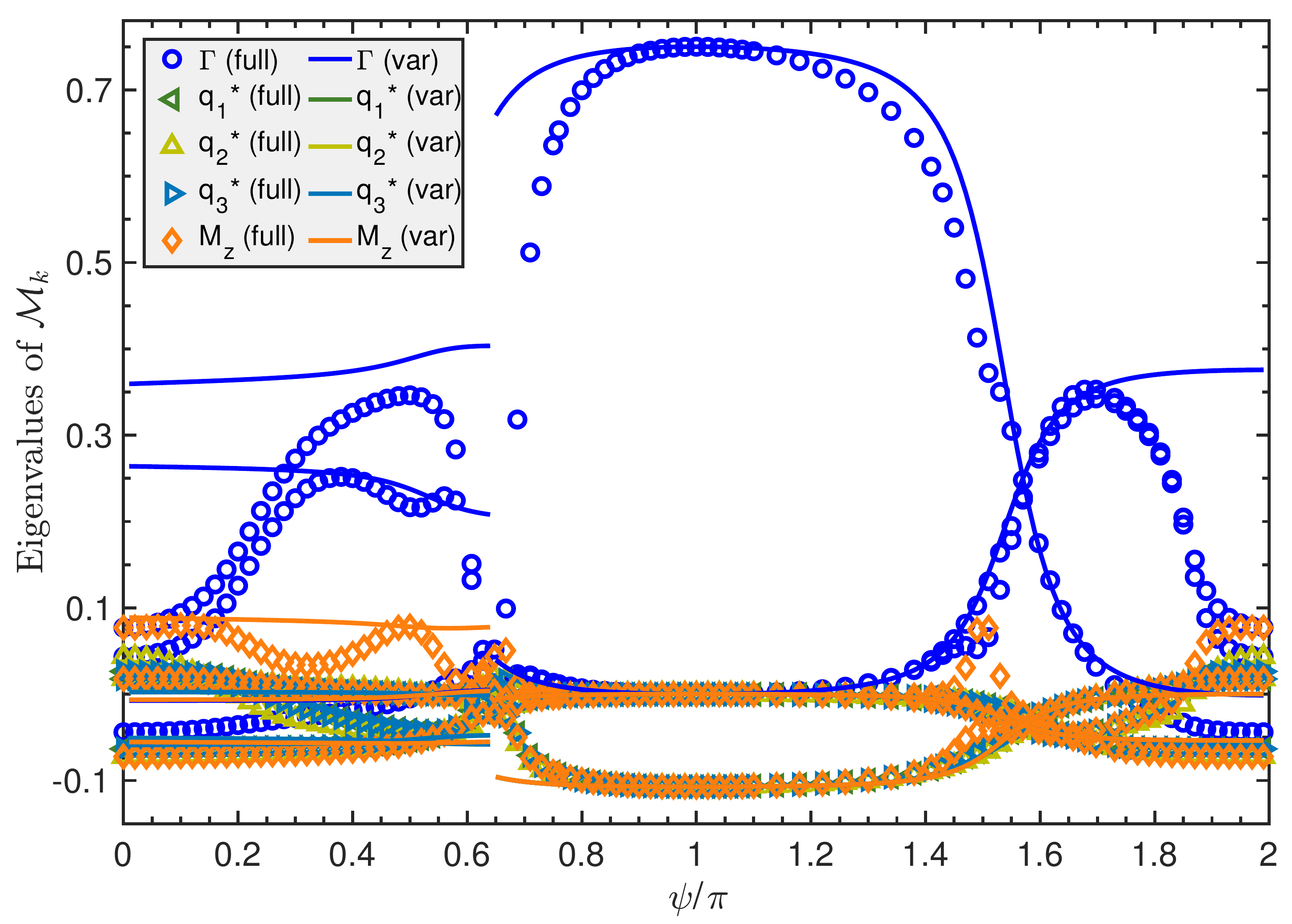}
\caption{Evolution of the eigenvalues of the correlation matrix $\mc{M}_{\bf k}$ of Eq.~(\ref{eq:CorMatrix}), evaluated in the GS of the 24-site cluster and for all allowed momenta of the cluster. There are two sets of data, one from ED in the full basis (symbols) and the other from ED in the orthonormalized classical basis of each region.}\label{fig:slength24}
\end{figure}

\section{Coupling matrices $\Lambda^{(\alpha)}_{\bf k}$ ($\alpha=x$, $y$, $z$)}\label{app:LambdaMatrices}
As mentioned in the main text, the $9\times9$ coupling matrix $\bs{\Lambda}_{\bf k}$ of the Luttinger-Tisza method is block diagonal and has the following general form 
\be\label{eq:lambda}
\bs{\Lambda}_{\bf k}=\left(
\begin{array}{ccc}
\Lambda^{(x)}_{\bf k}&0&0\\
0&\Lambda^{(y)}_{\bf k}&0\\
0&0&\Lambda^{(z)}_{\bf k}
\end{array}
\right),
\ee
with 
\bea
&&
\Lambda_{\bf k}^{(x)}=\frac{1}{2}
\left(\begin{array}{ccc}
0&J f_{1,\mathbf{k}}& J f_{2,\mathbf{k}}\\
J f_{1,-\mathbf{k}} &0&(J\!+\!K) f_{3,\mathbf{k}}\\
J f_{2,-\mathbf{k}}&(J\!+\!K) f_{3,-\mathbf{k}}&0
\end{array}\right),~\\
&&
\Lambda_{\bf k}^{(y)}=\frac{1}{2}
\left(\begin{array}{ccc}
0&J f_{1,\mathbf{k}}&(J\!+\!K) f_{2,\mathbf{k}}\\
J f_{1,-\mathbf{k}}&0&J f_{3,\mathbf{k}}\\
(J\!+\!K) f_{2,-\mathbf{k}} &J f_{3,-\mathbf{k}}&0
\end{array}\right),~\\
&&
\Lambda_{\bf k}^{(z)}=\frac{1}{2}
\left(\begin{array}{ccc}
0&(J\!+\!K) f_{1,\mathbf{k}}&J f_{2,\mathbf{k}}\\
(J\!+\!K) f_{1,-\mathbf{k}}&0&J f_{3,\mathbf{k}}\\
J f_{2,-\mathbf{k}} &J f_{3,-\mathbf{k}}&0
\end{array}\right),
\eea 
and 
$f_{\nu,\mathbf{k}}\!=\!1\!+\!e^{i\mathbf{k}\cdot\mathbf{a}_\nu}$, $\nu=1$-$3$, ${\bf a}_3\!=\!{\bf a}_2\!-\!{\bf a}_1$.  
By symmetry, the three matrices $\Lambda^{(\alpha)}_{\bf k}$ ($\alpha=x$, $y$, $z$) are related to each other by threefold rotations, namely 
\be
\Lambda^{(x)}_{\bf k}=\Lambda^{(y)}_{\mathsf{C}_3\cdot {\bf k}}=\Lambda^{(z)}_{\mathsf{C}_3^2\cdot {\bf k}}\,.
\ee  
As such, these matrices have the same overall spectrum of eigenvalues. 
The eigenvalues of $\Lambda^{(z)}_\mathbf{k}$ can be written in the following analytical form 
\be
\lambda_{\nu,{\bf k}}^{(z)}=2\sqrt{P_{\bf k}/3}\cos\left(\Phi_\mathbf{k}+\nu~2\pi/3\right)\,,\nu=1-3\\
\ee
where $P_{\bf k}$ and $\Phi_\mathbf{k}$ can be found via the relations
\be
\renewcommand{\arraystretch}{1.45}
\!\!\!\!\!\begin{array}{l}
\sqrt{P_{\bf k}/3}\cos(3\Phi_\mathbf{k})=3Q_\mathbf{k}/(2P_{\bf k}),\\
Q_\mathbf{k}\!=\!2J^2(J\!+\!K)\cos^2\left(\frac{{\bf k}\cdot{\bf a}_1}{2}\right) \cos\left(\frac{{\bf k}\cdot{\bf a}_2}{2}\right) \cos\left(\frac{{\bf k}\cdot {\bf a}_3}{2}\right)\,,\\
P_\mathbf{k}\!=\!(J\!+\!K)^2\cos^2\left(\frac{{\bf k}\cdot{\bf a}_1}{2}\right)\!+\!J^2\left[\cos^2\left(\frac{{\bf k}\cdot{\bf a}_3}{2}\right)\!+\!\cos^2\left(\frac{{\bf k}\cdot{\bf a}_2}{2}\right)\right].
\end{array}
\ee

\section{General setting of real space perturbation theory}\label{app:RSPT}
We remind here the general setting of the real space perturbation expansion~\cite{Lindgard1988,Long1989,Heinila1993,Mike2014,IoannisGammaModel,IoannisClassKitaev}. 
In this expansion one calculates the energy renormalization of a given classical GS from quantum fluctuations, in the following way. First we fix the classical GS, and parametrize the direction of each spin in that state by a local axis $\vec{e}_i^z$. We then choose two perpendicular axes $\vec{e}_i^x$ and $\vec{e}_i^y$ and define the combinations $\vec{e}_i^\pm=\frac{1}{2}(\vec{e}_i^x\pm i \vec{e}_i^y)$. Rewriting spin operators in the local frame, 
\bea
\vec{S}_i=S_i^z \vec{e}_i^z+S_i^+ \vec{e}_i^- + S_i^- \vec{e}_i^+\,.
\eea
gives the following general form of the Hamiltonian
\bea
\mc{H} &=& \frac{1}{2}\sum_{ij} \vec{S}_i\cdot\vec{A}_{ij}\cdot \vec{S}_j
=\frac{1}{2}\sum_{ij} \Big\{
A_{ij}^{zz} S_i^z S_j^z + \Big[
\Big(A_{ij}^{z+} S_i^z S_j^-  \nonumber\\
&+&
A_{ij}^{+z} S_i^- S_j^z 
+ A_{ij}^{++} S_i^- S_j^-
+ A_{ij}^{+-} S_i^- S_j^+ 
\Big) + h.c.\Big]
\Big\},
\eea
where $\bf{A}$ is a second-rank tensor containing the Heisenberg and Kitaev interactions with 
\be
A_{ij}^{zz}\!=\!\vec{e}_i^z\cdot\vec{A}_{ij}\cdot\vec{e}_j^z, ~~ A_{ij}^{z+}\!=\!\vec{e}_i^z\cdot\vec{A}_{ij}\cdot\vec{e}_j^+, ~~\text{etc}.
\ee
Next, we define the deviation operator $n_i \!=\! S-S_i^z$ and rewrite
\be
\renewcommand{\arraystretch}{1.25}
\begin{array}{l}
\mc{H}=E_{cl} + \sum_{j} B_j n_j + \frac{1}{2}\sum_{ij} A_{ij}^{zz} n_i n_j \\
+\frac{1}{2}\sum_{ij}\left( 
A_{ij}^{++}\!\!\!\!\!\!\!\!\underbrace{S_i^- S_j^-}_{\text{double spin-flip}} 
\!\!\!\!\!+~A_{ij}^{+-} \!\!\!\!\!\!\!\!\!\underbrace{S_i^- S_j^+}_{\text{spin-flip hopping}}
\!\!\!\!\!\!\!\!-2 A_{ij}^{z+} \!\!\!\!\!\underbrace{n_i S_j^-}_{\text{single spin-flip}} 
\!\!\!\!\!+~ h.c. \right)\,,
\end{array}
\ee
where $E_{cl}\!=\!S^2/2\sum_{ij} A_{ij}^{zz}$ is the classical energy, $\vec{B}_j\!=\!- S \sum_{i} \vec{e}_i^z \cdot \vec{A}_{ij} \!=\! B_j \vec{e}_j^z$ is the local exchange field on site $j$ (with magnitude $B_j\!=\!-S \sum_i A_{ij}^{zz}$), and we have also used the relation
\be
S \sum_{i}  A_{ij}^{z+} S_j^- = -B_j \vec{e}_j^z \cdot \vec{e}_j^+ = 0\,.
\ee
We then set 
\be\label{eq:H0V}
\renewcommand{\arraystretch}{1.25}
\begin{array}{l}
\mc{H}_0=E_{cl} + \sum_{j} B_j n_j + \frac{1}{2}\sum_{ij} A_{ij}^{zz} n_i n_j\,,\\
\mc{V}=\mc{H}-\mc{H}_0=\mc{V}_{1}+\mc{V}_{2}+\mc{V}_{3}\,,
\end{array}
\ee
where 
\be
\renewcommand{\arraystretch}{1.25}
\begin{array}{l}
\mc{V}_1 = \frac{1}{2}\sum_{ij} \left( A_{ij}^{++} S_i^- S_j^- + h.c. \right)\,,\\
\mc{V}_2 = \frac{1}{2}\sum_{ij} \left( A_{ij}^{+-}S_i^- S_j^+ + h.c. \right)\,,\\
\mc{V}_3 = -\sum_{ij}   \left( A_{ij}^{z+} n_i S_j^- + h.c. \right)\,,
\end{array}
\ee
stand for the the double spin-flip processes ($\mc{V}_1$ ), the single spin-flip hopping ($\mc{V}_2$), and the correlated, single spin-flip processes ($\mc{V}_3$). The latter are analogous to the cubic magnon terms in the standard Holstein-Primakof spin-wave expansion. 

Equation (\ref{eq:H0V}) form the basis for the RSPT, which proceeds via a perturbation theory in powers of $\mc{V}$. In the present problem, the leading contributions to the order by disorder effect arise in fourth order, as we discussed in Sec.~\ref{sec:ObD}.

\section{ED results for the 24-site cluster}\label{app:ED24site}
Figure~\ref{fig:spectrum24full} shows the low-energy spectrum of the 24-site cluster as obtained from ED in the full basis. Apart from certain unimportant details (related to the different symmetry sectors of this cluster), the results show the same qualitative picture with that from the 27-site cluster discussed in the main text. 
Similar qualitative agreement arises for the eigenvalues of the spin-spin correlation matrix of the 24-site cluster, which are shown in Fig.~\ref{fig:slength24}.

\bibliography{references}

\end{document}